\documentclass[12pt]{article}

\usepackage{scicite}
\usepackage{graphicx}
\usepackage{times}

\usepackage{amsmath,amssymb}
\usepackage{xcolor,soul}
\usepackage{gensymb}
\usepackage{float}
\newenvironment{sciabstract}{%
\begin{quote} \bf}
{\end{quote}}

\title{Emergent ferromagnetism near three-quarters filling\\ in twisted bilayer graphene} 

\author
{Aaron L. Sharpe,$^{1,2\ast}$ Eli J. Fox,$^{2,3\ast}$ Arthur W. Barnard,$^{3}$ Joe Finney,$^{3}$\\ Kenji Watanabe,$^{4}$  Takashi Taniguchi,$^{4}$\\ M. A. Kastner,$^{3,5,6}$  David Goldhaber-Gordon$^{2,3\dagger}$
\\
\normalsize{$^{1}$Department of Applied Physics, Stanford University,}\\
\normalsize{348 Via Pueblo Mall, Stanford, CA 94305, USA}\\
\normalsize{$^{2}$Stanford Institute for Materials and Energy Sciences,}\\
\normalsize{SLAC National Accelerator Laboratory,}\\
\normalsize{2575 Sand Hill Road, Menlo Park, California 94025, USA}\\
\normalsize{$^{3}$Department of Physics, Stanford University,}\\
\normalsize{382 Via Pueblo Mall, Stanford, CA 94305, USA}\\
\normalsize{$^{4}$National Institute for Materials Science,}\\
\normalsize{Namiki 1-1, Tsukuba, Ibaraki 305-0044, Japan}\\
\normalsize{$^{5}$Science Philanthropy Alliance,}\\
\normalsize{480 California Avenue \#304, Palo Alto, CA 94306, USA}\\
\normalsize{$^{6}$Department of Physics, Massachusetts Institute of Technology,}\\
\normalsize{77 Massachusetts Avenue, Cambridge, MA 02139, USA}\\
\\
\normalsize{$^\ast$These authors contributed equally to this work.}
\\
\normalsize{$^\dagger$To whom correspondence should be addressed;}\\ \normalsize{E-mail:  goldhaber-gordon@stanford.edu.}
}

\date{}
\begin{document}

\baselineskip24pt

\maketitle 

\begin{sciabstract}

    When two sheets of graphene are stacked at a small twist angle, the resulting flat superlattice minibands are expected to strongly enhance electron-electron interactions. Here we present evidence that near three-quarters ($3/4$) filling of the conduction miniband these enhanced interactions drive the twisted bilayer graphene into a ferromagnetic state. We observe emergent ferromagnetic hysteresis, with a giant anomalous Hall (AH) effect as large as $10.4\ \mathrm{k\Omega}$ and signs of chiral edge states in a narrow density range around an apparent insulating state at $3/4$. Surprisingly, the magnetization of the sample can be reversed by applying a small DC current. Although the AH resistance is not quantized and dissipation is significant, we suggest that the system is an incipient Chern insulator.

\end{sciabstract}


In weakly dispersing bands, electron-electron interactions dominate over kinetic energy, often leading to interesting correlated phases. Graphene has emerged as a preeminent platform for investigating such flat bands because of the control of the band structure enabled by stacking multiple layers and the tunability of the band filling via electrostatic gating. In particular, the moir\'e superlattice of so-called ``magic-angle'' twisted bilayer graphene (TBG), in which one monolayer graphene sheet is stacked on top of another with a relative angle of rotation between the two crystal lattices of near one degree, is predicted to host nearly flat bands of ${\sim} 10\ \mathrm{meV}$ width~\cite{Bistritzer2011,Fang2016,Nam2017}.

In the single-particle picture, the flat bands are four-fold degenerate because of spin and valley symmetries~\cite{Cao2016}. However, magic-angle TBG has recently been shown to exhibit high-resistance states at half ($1/2$) and three-quarter ($3/4$) filling of the conduction and valence bands~\cite{Cao2018a,Yankowitz2018} and at one-quarter ($1/4$) filling of the conduction band~\cite{Yankowitz2018}, all cases where metallic behavior would be expected in the absence of interactions. Surprisingly, magic-angle TBG can become superconducting when doping slightly away from $1/2$ filling of either the conduction or valence band~\cite{Yankowitz2018,Cao2018b}.

Theoretical calculations have raised the possibility of magnetic order as a result of interactions lifting spin and valley degeneracies~\cite{Xie2018,Ochi2018,Dodaro2018,Thomson2018}. Here, we present unambiguous experimental evidence of emergent ferromagnetism at $3/4$ filling of the conduction band in TBG: a giant anomalous Hall (AH) effect that displays hysteresis in magnetic field. We also find evidence of chiral edge conduction. Our results suggest that the $3/4$ filling state is a correlated Chern insulator.


We used a ``tear-and-stack'' dry-transfer method~\cite{Cao2016,Kim2016} and standard lithographic techniques to fabricate a TBG Hall bar device (Fig.~\ref{fig:fig1}, inset) with a target twist angle $\theta = 1.17^\circ$. The graphene is encapsulated in two hexagonal boron nitride (hBN) cladding layers to protect the channel from disorder and to act as dielectrics for electrostatic gating. With both a silicon back gate and Ti/Au top gate, we can independently tune the charge density $n$ in the TBG and the perpendicular displacement field $D$~\cite{Sharpe2018Supplement,Oostinga2008}.

We measured the longitudinal and Hall resistances using standard lock-in techniques with a $5\ \mathrm{nA}$ RMS AC bias current. A complicated electronic structure is revealed by the behavior of the longitudinal resistance as a function of $n$ and $D$ (Fig.~\ref{fig:fig1}, upper panel). We observe strong resistance peaks at the charge neutrality point (CNP) (identified from Landau fan diagrams~\cite{Sharpe2018Supplement}) and at densities $\pm n_s = 3.37\times 10^{12}\ \mathrm{cm}^{-2}$ corresponding to full filling of the mini-Brillouin zone (mBZ) of the TBG superlattice, with four electrons (or holes) per superlattice unit cell. This value of $n_s$ is consistent with a twist angle $\theta = 1.20^\circ \pm 0.01^\circ$ in the TBG heterostructure~\cite{Hunt2013}, very near our target angle of $1.17^\circ$. A slight kink in the positions of the CNP and other features as a function of displacement field is noticeable but not repeatable between cool-downs~\cite{Sharpe2018Supplement}.

Beyond the peaks expected from a single-particle picture of the TBG band structure, we observe additional high resistance states at $1/4$, $1/2$, and $3/4$ fillings of the mBZ. These fillings, corresponding to two and three electrons per superlattice unit cell, respectively, have previously been attributed to correlated insulating states~\cite{Yankowitz2018,Cao2018b}. Another unexpected peak at $n/n_s = -1.15$ and a corresponding shoulder on the full filling peak of the electron side (seen in Fig.~\ref{fig:fig1}, lower panel) do not correspond to expectations for TBG alone.  They likely result from the lattice alignment of the top graphene sheet with the top hBN layer, with the density $1.15 n_s$ corresponding to an angle $\theta = 0.81^\circ \pm 0.02^\circ$~\cite{Hunt2013}. Such near-alignment with the top hBN layer is borne out by optical images of the heterostructure; the bottom hBN is far from aligned with the bottom graphene sheet. This vertical asymmetry in the heterostructure may play a role in the strong dependence of the peak structure on the sign and magnitude of the displacement field~\cite{Sharpe2018Supplement}.


Magnetotransport in graphene-based heterostructures typically does not depend on the history of the applied field. Surprisingly, we find that in a narrow range of $n$ near $n/n_s=3/4$, transport is hysteretic with respect to an applied out-of-plane magnetic field $B$ (Fig.~\ref{fig:fig2}A).
When the applied field is swept to zero from a large negative value, a large AH resistance $R_{yx} \approx \pm 6$~k$\Omega$ remains, with the sign depending on the direction of the field sweep, indicating that the sample has a remanent magnetization. 
This large AH signal is especially striking given the absence of both transition metals (typically associated with magnetism) and heavy elements (to give spin-orbit coupling) in TBG. If the field is left at zero, the magnetization is very stable, with no significant change in the Hall resistance observed over the course of six hours~\cite{Sharpe2018Supplement}.  As the field is increased beyond a coercive field of order $100\ \mathrm{mT}$ opposite to the direction of the training field, the Hall signal changes sign, pointing to a reversal of the magnetization.

Multiple intermediate jumps appear near the coercive field; these are very repeatable over successive hysteresis loops~\cite{Sharpe2018Supplement} and likely correspond to either a mixed domain structure with varying coercivities or a repeatable pattern of domain wall motion and pinning. This behavior may result from inhomogeneity caused by local variations in the twist angle between the graphene sheets, which has recently been directly imaged using transmission electron microscopy~\cite{Yoo2018}, or by local variations in electrostatic potential~\cite{Xue2011}.

Hysteresis loops of $R_{yx}$ and $R_{xx}$ would ideally be antisymmetric and symmetric, respectively (in the sense that $R_{ij}(B) = \pm\widetilde{R}_{ij}(-B)$, where $R_{ij}$ and $\widetilde{R}_{ij}$ are measured with the field sweeping in opposite directions). We find that $R_{yx}$ hysteresis loops are roughly antisymmetric but offset vertically by $-1\ \mathrm{k\Omega}$. $R_{xx}$ is nearly flat with field, but has an antisymmetric component, presumably because of mixing in of the large changes in $R_{yx}$.

We define the coercive field as half the difference between the fields where the largest jumps in $R_{yx}$ occur on the upward and downward sweeps. With increasing temperature $T$, the coercive field steadily decreases before vanishing at $3.9\ \mathrm{K}$ (Fig.~\ref{fig:fig2}, C and D). This monotonic dependence is to be expected, since flipping individual domains or moving domain walls in a magnet is usually thermally activated~\cite{Emori2015}.

The Hall signal appears to be the sum of two parts: an anomalous component that reflects the sample magnetization~\cite{Nagaosa2010}, and a conventional component linear in field with a Hall slope $R_\text{H}$ (Fig.~\ref{fig:fig2}B; see the supplementary text for how we separate these two components). Unlike the coercive field, the magnitude of the residual anomalous Hall resistance at zero field, which we denote by $R_{yx}^\text{AH}$, does not vary monotonically with temperature: $R_{yx}^\text{AH}$ rises slightly with increasing $T$ up to $2.8\ \mathrm{K}$, before rapidly falling to zero by $5\ \mathrm{K}$ (Fig.~\ref{fig:fig2}, C and D).

Although the hysteresis is observable over a wide range of displacement fields~\cite{Sharpe2018Supplement}, it only emerges in a narrow range of densities near $3/4$ filling of the mBZ. $R_{yx}^\text{AH}$ displays a sharp peak as a function of $n/n_s$, reaching $6.6\ \mathrm{k\Omega}$ for $n/n_s = 0.758$ with a full width at half maximum of $0.04 n_s$ (Fig.~\ref{fig:fig2}B). These measurements were made along a trajectory for which $D$ changes by approximately $10\%$ coincident with the primary intended change in $n$~\cite{Sharpe2018Supplement}. In a separate measurement, we observed loops with $R_{yx}^\text{AH}$ up to $10.4\ \mathrm{k\Omega}$~\cite{Sharpe2018Supplement}.

The gate-voltage dependence of the conventional linear Hall slope $R_H$~\cite{Sharpe2018Supplement} appears typical for a transition from $p$-type- to $n$-type-dominated conduction in a semimetal or small-gap semiconductor, with $\left|R_H\right|$ rising when approaching the transition from either side, then turning over and crossing through zero (Fig.~\ref{fig:fig2}B). Recent studies of near-magic-angle TBG have reported high resistance at $3/4$ filling~\cite{Yankowitz2018,Cao2018b} (cf. our Fig.~\ref{fig:fig1}), suggesting that spin and valley symmetries are spontaneously broken, resulting in a low density of states (or a gap) at this filling. Our results similarly indicate a possible correlated insulating state, here with an AH effect in a narrow range of densities around this same filling.


The presence of a giant AH effect in an apparent insulator is reminiscent of a ferromagnetic topological insulator approaching a Chern insulator state~\cite{Chang2013b,Checkelsky2014,Kou2014}, where they would exhibit a quantum anomalous Hall (QAH) effect: longitudinal resistivity $\rho_{xx}$ approaches zero and Hall resistivity $\rho_{yx}$ is quantized to $h/Ce^2$~\cite{Fox2018,Yu2010}, where $h$ is Planck's constant, $e$ is the electron charge, and $C$ is the Chern number arising from the Berry curvature of the filled bands ($C = \pm 1$ in presently available QAH materials).
Chiral edge modes associated with a quantized Hall system manifest in nonlocal transport measurements~\cite{Bestwick2015,Chang2015}. In an ideal QAH system described by the B\"uttiker edge state model~\cite{Buttiker1988}, floating metallic contacts equilibrate with the chiral edge states that propagate into them. Clearly our results are not those of an ideal QAH system. Dissipation can cause deviations from the ideal behavior, while still giving results differing from classical diffusive transport. Below we present and analyze our experimental evidence for nonlocal transport in the magnetic state.

The three-terminal resistance $R_{54,14}$, where $R_{ij,k\ell} = V_{k \ell}/I_{ij}$ with $V_{k\ell}$ the voltage between terminals $k$ and $\ell$ when a current $I_{ij}$ flows from terminal $i$ to $j$, is shown in Fig.~\ref{fig:fig3} (upper panel) for two values of $n/n_s$. When the density is tuned away from the center of the magnetic regime, $R_{54,14}$ is ${\sim}5\ \mathrm{k\Omega}$ and nearly independent of applied field. We ascribe this behavior to diffusive bulk transport and a finite contact resistance to ground. By contrast, at the center of the magnetic regime we observe a hysteresis loop with $R^\downarrow_{54,14}= 3.3\ \mathrm{k\Omega}$ and $R^\uparrow_{54,14}= 9.1\ \mathrm{k\Omega}$, where $R^{\uparrow (\downarrow)}_{ij,k\ell}$ are the remanent resistances as defined above. The difference $|R^\uparrow_{54,14}-R^\downarrow_{54,14}|$ is largest near the peak in $R_{yx}^\text{AH}$ shown in Fig.~\ref{fig:fig2}B. For a QAH effect, we would expect $R_{54,14}$ to be either 0 or $h/Ce^2$ ($25{,}813 \ \mathrm{\Omega}$ for $C=1$). While the difference $|R^\uparrow_{54,14}-R^\downarrow_{54,14}| = 5.8\ \mathrm{k\Omega}$ is smaller than the ideal $C=1$ QAH case by a factor of $4$, it could be consistent with a QAH state in combination with other dissipative transport mechanisms or a complex network of domain walls (in addition to contact resistance). These three-terminal measurements alone cannot rule out diffusive bulk transport with a very large (anomalous) Hall coefficient, but four-terminal measurements suggest this is unlikely.

In contrast to the three-terminal case, four-terminal nonlocal resistances where the voltage is measured far from the current path are exponentially small in the case of homogeneous diffusive conduction~\cite{vanderPauw1958}. For $n/n_s = 0.725$, away from the peak in $R_{yx}^\text{AH}$, the measured $R_{45,12}= 10\ \mathrm{\Omega}$ (Fig.~\ref{fig:fig3}, lower panel) is indeed small. In the magnetic regime at $n/n_s = 0.749$, however, the four-terminal resistance is two orders of magnitude larger than the $3\ \mathrm{\Omega}$ expected from homogeneous bulk conduction, with a hysteresis loop yielding $R^\downarrow_{54,12}= 42\ \mathrm{\Omega}$ and $R^\uparrow_{54,12}= 240\ \mathrm{\Omega}$. Although in an ideal QAH state with pure chiral edge conduction this four-terminal resistance would be zero, the presence of additional conduction paths, such as extra non-chiral edge states~\cite{Wang2013a}, parallel bulk conduction, or transport along magnetic domain walls~\cite{Rosen2017,Yasuda2017}, can result in large, hysteretic nonlocal resistances. 


Surprisingly, we find that the $n/n_s = 3/4$ state is extremely sensitive to an applied DC current. All of the measurements described above were performed with a 5~nA RMS AC bias current, but we observed curious behavior when we added a DC bias $I_\text{DC}$ to this small AC signal. Sweeping $I_\text{DC}$ between $\pm 50$~nA with $B = 0$ (Fig.~\ref{fig:fig4}), we found that the differential Hall resistance $dV_{yx}/dI$ follows a hysteresis loop reminiscent of its magnetic field dependence. This loop was very repeatable after a slight deviation from the first trace (black trace, Fig.~\ref{fig:fig4}), for which $I_\text{DC}$ was ramped from 0 to $-50$~nA after first magnetizing the sample in a $-500$~mT field. Additional details about the nature of the jumps in differential resistance and the effect of external magnetic field on the hysteresis loops are presented in the supplemental text.

The switching of $dV_{yx}/dI$ is clear evidence that, like the external magnetic field, the applied DC current bias modifies the magnetization. This phenomenon might be similar to switching in other ferromagnetic materials, in which spin-transfer or spin-orbit torques can influence the magnetization. However, the current necessary to flip the moment appears to be very small~\cite{Apalkov2016}. It has also been proposed that a current could efficiently drive domain wall motion in a QAH system due to quantum interference effects from the edge states~\cite{Upadhyaya2016}.


Our observation of a large hysteretic AH effect establishes a ferromagnetic moment associated with the apparent $3/4$ correlated insulating state. Specifically, we suggest that this state is a Chern insulator. Extrinsic mechanisms for AH, based on scattering rather than band topology, cannot contribute to the Hall resistance of an insulator~\cite{Nagaosa2010}, yet the measured $R_{yx}^\text{AH}$ is largest at an apparent insulating state. Furthermore, our measurements yield a Hall angle $\rho_{yx}/\rho_{xx}$ up to 1.4, almost an order of magnitude larger than any other reported AH~\cite{Liu2018} apart from magnetic topological insulators exhibiting a QAH effect (here we convert our measured resistances to resistivities which we approximate as spatially homogeneous). With $\rho_{yx} \lesssim 0.4h/e^2$ and $\rho_{xx}\approx 0.3h/e^2$ the present device is clearly not an ideal Chern insulator. Yet after early magnetically doped topological insulators showed comparable values~\cite{Checkelsky2012,Chang2013a,Kou2013}, growth improvements in those materials soon yielded QAH~\cite{Chang2013b,Checkelsky2014,Kou2014}. If the present device is a nascent Chern insulator, the largest measured $R_{yx}^\text{AH}\approx h/2.5e^2$ limits the possible Chern number to $C = 1$ or $2$.

In combination with nonlocal transport that appears incompatible with homogeneous bulk conduction, the sheer magnitudes of the Hall and longitudinal resistances suggest a picture of chiral edge modes in combination with a poorly conducting bulk or a network of magnetic domain walls resulting from inhomogeneity (see the supplementary text for additional discussion). These possibilities can be directly explored in future experiments using spatially resolved magnetometry to search for domains, and transport in a Corbino geometry to measure bulk conduction independent of chiral edge modes if domain walls can be removed.

Achieving a Chern insulator state by definition requires opening a topologically nontrivial gap. The low energy flat minibands in magic-angle TBG are empirically isolated from higher order bands~\cite{Cao2016}, which is expected when taking into account mutual relaxation of the two layers' lattices~\cite{Nam2017}. The low energy conductance and valence minibands have been variously predicted to meet at Dirac points at the CNP, which may~\cite{Po2018,Zou2018} or may not~\cite{Kang2018,Koshino2018} be symmetry protected. The rotational alignment of the TBG to one of the hBN cladding layers in our device could thus be key to the observed AH effect: the associated periodic moir\'e potential should on average break A-B sublattice symmetry, opening or enhancing a gap at the mini-Dirac points. A gap associated with such symmetry breaking has been seen~\cite{Amet2013,AmetThesis,Hunt2013} and explained~\cite{Moon2014,Mucha-Kruczynski2013,Jung2015} in heterostructures of monolayer or Bernal-stacked bilayer graphene with hBN. At $3/4$ filling of the conduction band of thus-gapped magic angle TBG, spin and valley symmetry may be spontaneously broken, and 3 of the 4 flavors filled with the other empty. This scenario could account for our observation of an apparent Chern insulator. Indeed, Ref.~\citen{Xie2018} predicts a QAH effect arising in TBG (without aligned hBN) at $3/4$ filling from such a mechanism (see Ref.~\citen{Zhang2018} for a prediction of a similar situation in graphene-based moir\'e systems).

Aside from the topological aspect, the appearance of magnetism in this system is striking. Unlike previous studies of graphitic carbon exhibiting magnetism due to defects~\cite{Esquinazi2003,Mombru2005,Pardo2006,Cervenka2009} or adsorbed impurities~\cite{Yeh1989,Nicholls1990,Gao2015}, the order in the present device appears to emerge because of interactions in a clean graphene-based system; the anomalous Hall signal appears only in a narrow range of densities around a state that may be spin and valley polarized. Such intrinsic magnetism also stands in contrast to the magnetic topological insulators, where exchange coupling is induced through doping with transition metals~\cite{Yu2010,Chang2013b,Chang2015}. Further experiment and theory will be needed to elucidate the order parameter, which may have both spin and orbital components or break spatial symmetry (cf. Ref.~\citen{Zhou2016} for a model in which an antiferromagnet is a Chern insulator).

The discovery of a possible new platform for QAH physics, less disordered than the familiar magnetic chalcogenide alloys, may offer hope for more robust quantization, with applications in metrology~\cite{Fox2018}, quantum computation~\cite{Lian2018,He2017,Mahoney2017}, or low-power-consumption electronics. The ability to switch the magnetization in TBG with an applied DC current might have practical applications in extremely low-power magnetic memory architectures, given the orders-of-magnitude smaller critical current density required for flipping the magnetization compared to prior devices~\cite{Apalkov2016}. More broadly, understanding the magnetic order and topological character of the correlated insulating states will be crucial to unraveling the rich phase diagram of TBG.



\section*{Acknowledgments}
We acknowledge fruitful discussions with Michael Zaletel, Allan MacDonald, Ming Xie, Todadri Senthil, Steve Kivelson, Yoni Schattner, Nick Bultinck, Patrick Gallagher, Feng Wang, Matt Yankowitz, and Guorui Chen. Yuan Cao and Pablo Jarillo-Herrero generously taught us about their fab process and their insights into TBG. Hava Schwartz and Sungyeon Yang helped with device fabrication, and they and Anthony Chen performed preliminary measurements as part of a project-based lab class at Stanford. Device fabrication, measurements, and analysis were supported by the U.S.\ Department of Energy, Office of Science, Basic Energy Sciences, Materials Sciences and Engineering Division, under Contract DE-AC02-76SF00515. Infrastructure and cryostat support were funded in part by the Gordon and Betty Moore Foundation through Grant GBMF3429. Part of this work was performed at the Stanford Nano Shared Facilities (SNSF), supported by the National Science Foundation under award ECCS-1542152. A.~L.~S. acknowledges support from a Ford Foundation Predoctoral Fellowship and a National Science Foundation Graduate Research Fellowship. E.~F. acknowledges support from an ARCS Foundation Fellowship. K.W. and T.T. acknowledge support from the Elemental Strategy Initiative conducted by the MEXT, Japan and and the CREST (JPMJCR15F3), JST.

\section*{Figures}
\begin{figure}
    \centering
    \includegraphics[width=12cm]{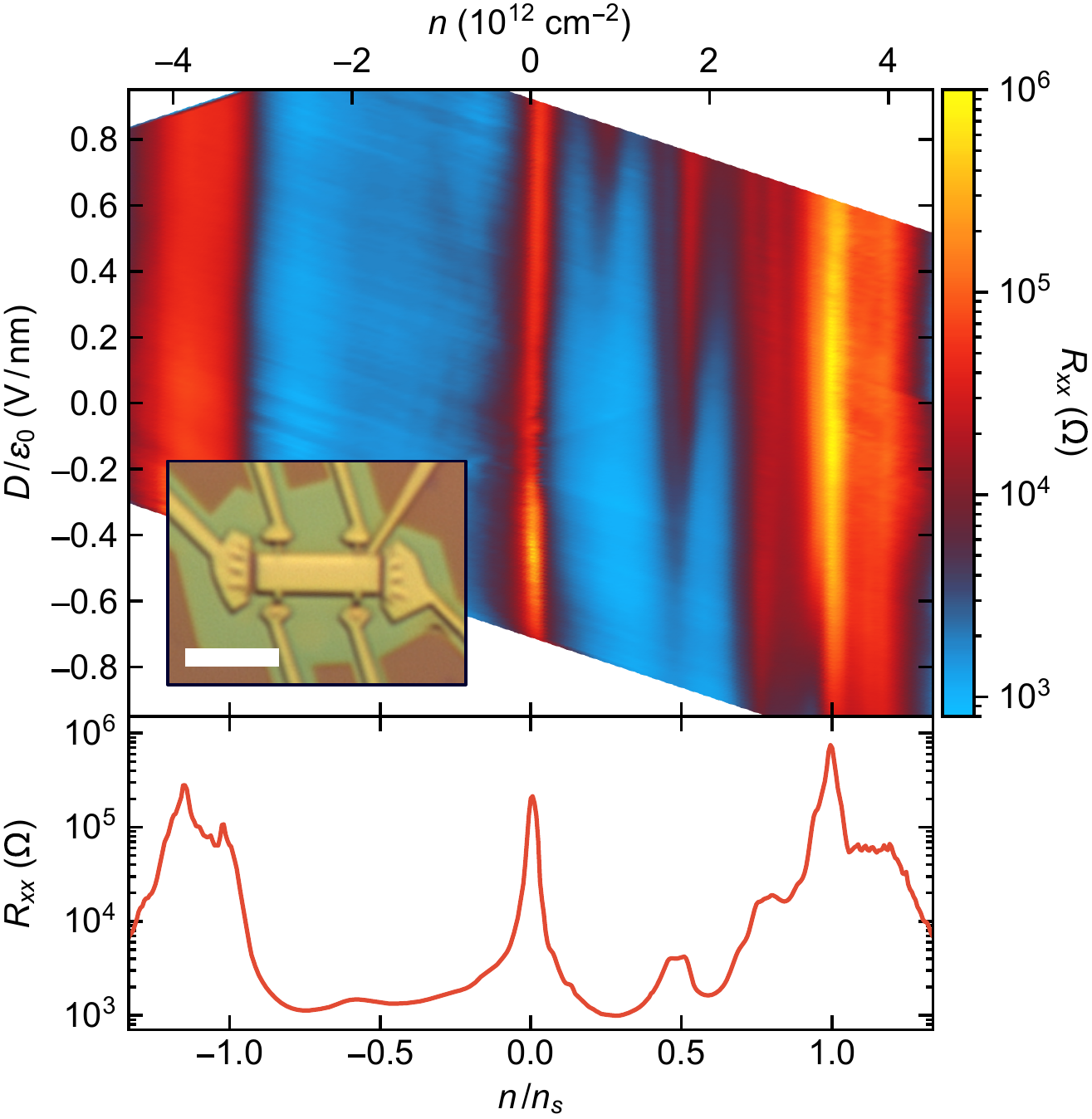}
    \caption{\textbf{Correlated states in near-magic-angle TBG.} (Upper panel) Longitudinal resistance $R_{xx}$ of the TBG device (measured between contacts separated by $2.15$ squares) as a function of carrier density $n$ (shown on the top axis) and perpendicular displacement field $D$ (left axis), which are tuned by the top- and back-gate voltages, at $2.1\ \mathrm{K}$.
    $n$ is mapped to a filling factor relative to the superlattice density $n_s$, corresponding to four electrons per moir\'e unit cell, shown on the bottom axis.
    (Inset) Optical micrograph of the completed device. The scale bar is $5\ \mathrm{\mu m}$.
    (Lower panel) Line cut of $R_{xx}$ with respect to $n$ taken at $D/\epsilon_0 = -0.22\ \mathrm{V/nm}$ showing the resistance peaks at full filling of the superlattice, and additional peaks likely corresponding to correlated states emerging at intermediate fillings.}
    \label{fig:fig1}
\end{figure}

\begin{figure}
    \centering
    \includegraphics[width=12cm]{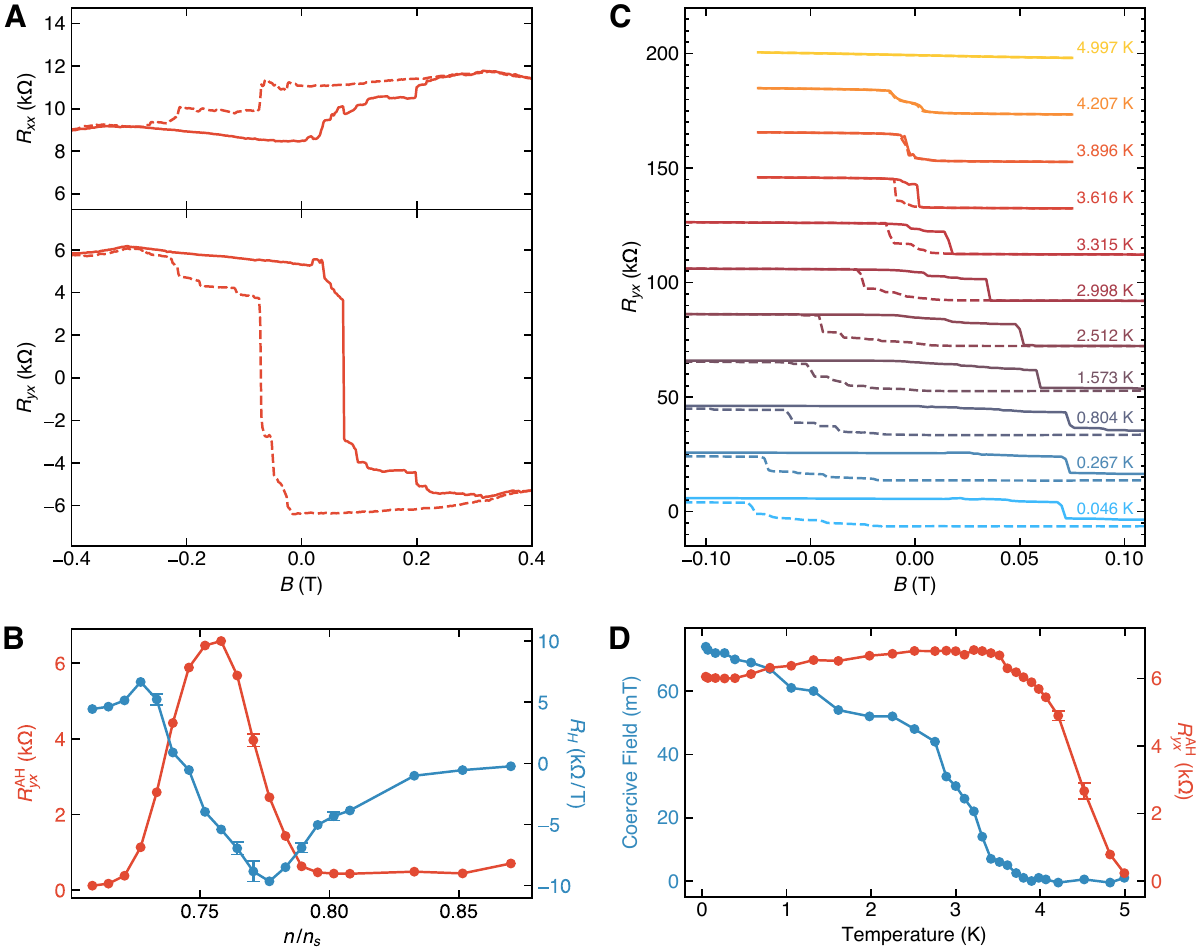}
    \caption{\textbf{Emergent ferromagnetism near three-quarters filling.} (\textbf{A}) Magnetic field dependence of the longitudinal resistance $R_{xx}$ (upper panel) and Hall resistance $R_{yx}$ (lower panel) with $n/n_s=0.746$ and $D/\epsilon_0 = -0.62\ \mathrm{V/nm}$ at $30\ \mathrm{mK}$, demonstrating a hysteretic anomalous Hall effect resulting from emergent magnetic order.
    The solid and dashed lines correspond to measurements taken while sweeping the magnetic field $B$ up and down, respectively.
    (\textbf{B}) Zero-field anomalous Hall resistance $R_{yx}^\text{AH}$ (red) and ordinary Hall slope $R_H$ (blue) as a function of $n/n_s$ for $D \approx -0.6\ \mathrm{V/nm}$. $R_{yx}^\text{AH}$ is peaked sharply with a maximum around $n/n_s = 0.758$, coincident with $R_H$ changing sign. These parameters are extracted from line fits of $R_{yx}$ versus $B$ on the upward and downward sweeping traces in a region where the $B$-dependence appears dominated by the ordinary Hall effect.
    The error bars reflect fitting parameter uncertainty along with the effect of varying the fitting window, and are omitted when smaller than the marker. 
    (\textbf{C}) Temperature dependence of $R_{yx}$ versus $B$ at $D\epsilon_0 = -0.62\ \mathrm{V/nm}$ and $n/n_s = n/n_s=0.746$ between $46\ \mathrm{mK}$ and $5.0\ \mathrm{K}$, showing the hysteresis loop closing with increasing temperature. Successive curves are offset vertically by $20\ \mathrm{k\Omega}$ for clarity. 
    (\textbf{D}) Coercive field and anomalous Hall resistance (extracted using the same fitting procedure as above) plotted as a function of temperature from the same data partially shown in (C).
    Data in Fig.~\ref{fig:fig2} were taken during a separate cooldown from that of the data in the rest of the figures, but show representative behavior~\cite{Sharpe2018Supplement}.}
    \label{fig:fig2}
\end{figure}

\begin{figure}
    \centering
    \includegraphics[width=12cm]{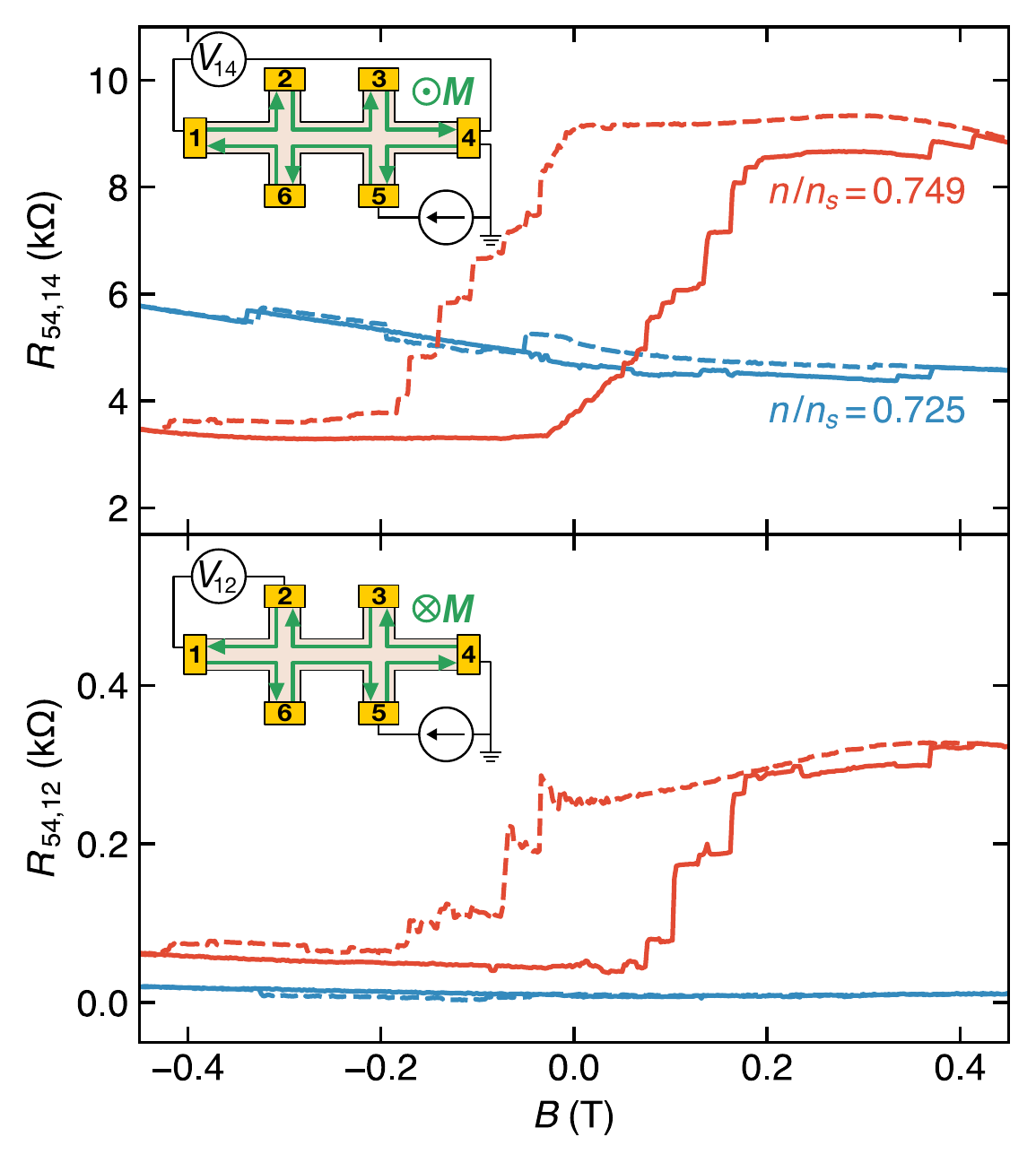}
    \caption{\textbf{Nonlocal resistances providing evidence of chiral edge states.} 
    Three- and four- terminal nonlocal resistances $R_{54,14}$ and $R_{54,12}$, measured at $2.1\ \mathrm{K}$ with $D/\epsilon_0 = -0.22\ \mathrm{V/nm}$, are shown in the upper and lower panels, respectively. For $n/n_s = 0.725$ (blue) away from the peak in AH resistance $R_{yx}^\text{AH}$, the nonlocal resistances are consistent with diffusive bulk transport. However, with $n/n_s = 0.749$ (red) in the magnetic regime where $R_{yx}^\text{AH}$ is maximal, large, hysteretic nonlocal resistances suggest chiral edge states are present.
    The inset schematics display the respective measurement configurations. Green arrows in the upper inset represent the apparent edge state chirality for positive magnetization, while in the lower inset they reflect negative magnetization.}
    \label{fig:fig3}
\end{figure}

\begin{figure}
    \centering
    \includegraphics[width=12cm]{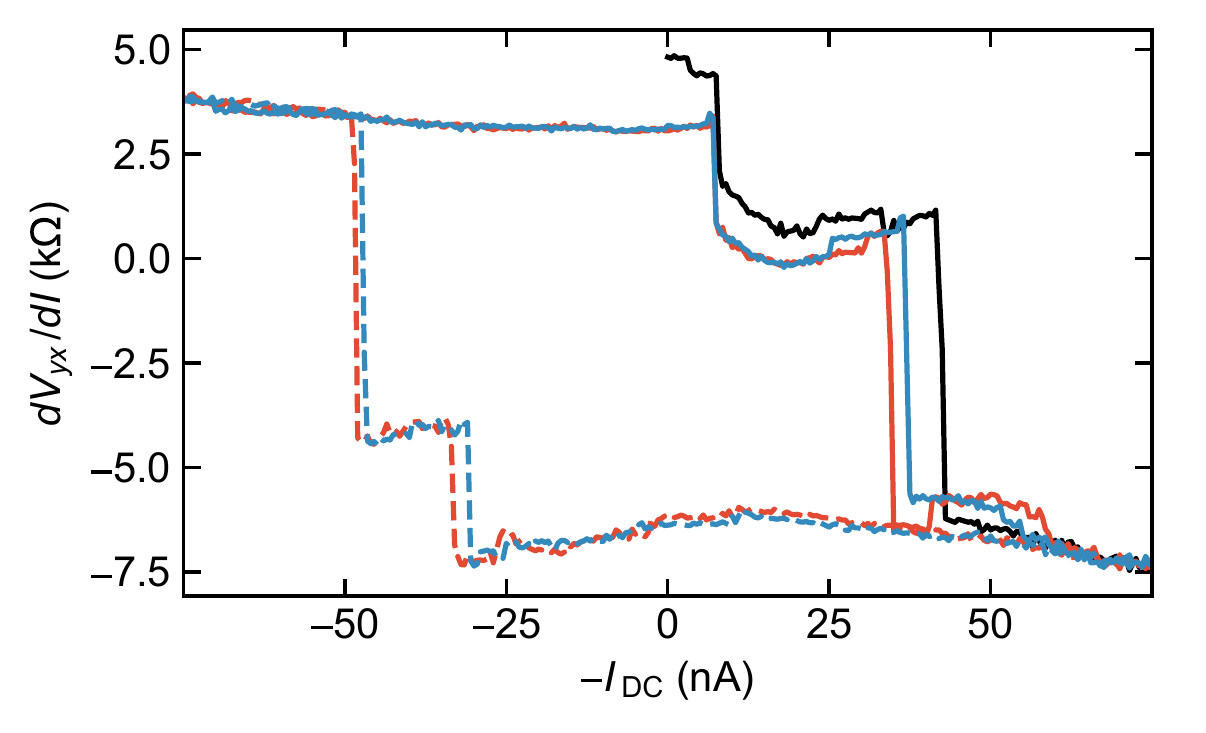}
    \caption{\textbf{Current-driven switching of the magnetization.}
    Differential Hall resistance $dV_{yx}/dI$ measured with a $5\ \mathrm{nA}$ AC bias as a function of an applied DC current $I_\text{DC}$ at $2.1\ \mathrm{K}$ with $D/\epsilon_0 = -0.22\ \mathrm{V/nm}$ and $n/n_s = 0.749$.
    After magnetizing the sample in a $-500\ \mathrm{mT}$ field and returning to $B=0$, $I_\text{DC}$ was swept from $0$ to $-50\ \mathrm{nA}$ (black trace), resulting in $dV_{yx}/dI$ changing sign. Successive loops in $I_\text{DC}$ between $\pm 50\ \mathrm{nA}$ demonstrate reversible and repeatable switching of the differential Hall resistance (red and blue, with solid and dashed traces corresponding to opposite sweep directions).
    Note that $dV_{yx}/dI$ is plotted against $-I_\text{DC}$ for better comparison with magnetic field hysteresis loops.}
    \label{fig:fig4}
\end{figure}

\clearpage

\section*{Supplementary Materials}

\setcounter{section}{3}
\setcounter{equation}{0}
\setcounter{figure}{0}
\renewcommand{\theequation}{S\arabic{equation}}
\renewcommand{\thefigure}{S\arabic{figure}}

\baselineskip18pt

\subsection*{Methods}

Our device consists of twisted bilayer graphene (TBG) fully encapsulated in two hexagonal boron nitride (hBN) cladding layers, each ${\sim} 50\ \mathrm{nm}$ thick. The heterostructure was assembled using a a Poly(Bisphenol A carbonate) film/gel (Gel-Pak DGL-17-X8) stamp on a glass slide heated to $60\ ^\circ\mathrm{C}$~\cite{Cao2016,Bhandari2016}. The stamp was first used to pick up the top hBN flake. Then, to stack two layers of graphene at a well defined twist angle, we used the van der Waals attraction between hBN and an exfoliated monolayer graphene flake to tear and pick up a portion of monolayer graphene from the larger flake. We then controllably rotated and picked up the remaining portion of the monolayer graphene ~\cite{Cao2016,Kim2016}. This process allows us to stack two layers of monolayer graphene to within $\pm 0.2^\circ$, with the precision limited by motion of the flake during the tearing process. The completed stack was transferred onto $300$-$\mathrm{nm}$-thick SiO$_2$ atop a degenerately doped Si substrate which is used as a back gate.

A Ti/Au top gate was deposited onto the completed heterostructure, and was subsequently used as a hard mask for a CHF$_3$/O$_2$ (50/5 sccm) etch to define a Hall bar region. During the mesa etch, the heterostructure was protected by resist extending outward from the hard mask near each of the leads of the Hall bar to provide space for making Cr/Au edge contacts~\cite{Wang2013b} without risk of shorting to the top gate. Throughout all processing, the sample temperature was kept below $180\ ^\circ\mathrm{C}$, to prevent potential relaxation of the twist angle of the TBG.

Using the Au top gate and the Si back gate, we can tune both the charge density in the TBG and the displacement field applied to the device. We model our gates as parallel plate capacitors such that the density under the top gated region is given by 
\[
n = C_{BG}(V_{BG}-V_{BG}^0) + C_{TG}(V_{TG}-V_{TG}^0),
\]
where $\mathrm{BG}$ ($\mathrm{TG}$) indicates the back (top) gate, $C$ is the capacitance per unit area determined from low-density Hall slope measurements, and $(V_{BG}^0,V_{TG}^0)$ is the charge neutrality point (CNP) of the top gated region at zero displacement field. We define the applied displacement field as
\[
D = (D_{BG} - D_{TG})/2,
\]
where the displacement field within a given dielectric $D_i = \epsilon_i (V_i - V_i^0)/d_i$, $\epsilon_i$ is the relative dielectric constant, and $d_i$ is the thickness of each dielectric. We assume a relative dielectric constant of $\epsilon_{TG}=3$ for hBN. As we do not see any clear features to ascribe to a true zero in displacement field, we assume that when both gates are at $0\ \mathrm{V}$, $D\approx 0$. This is a reasonable assumption given that the expected displacement field due to the work function difference between the top and back gate is small ($-0.01\ \mathrm{V/nm}$). We expect any nonzero displacement field at zero gate voltages to simply yield a constant offset to our reported values of displacement field.

The device was measured in a dilution refrigerator reaching a base temperature of $30\ \mathrm{mK}$. To obtain a low electron temperature in the device and to reduce high-frequency noise, the measurement lines are equipped with electronic filtering at the mixing chamber stage. Passing the wires through a cured mixture of epoxy and bronze powder filters GHz frequencies, while MHz frequencies are attenuated by low-pass RC filters mounted on sapphire plates for thermal anchoring. Four-terminal resistance measurements were performed with Stanford Research Systems SR830 lock-in amplifiers with NF Corporation LI-75A voltage preamplifiers, using a $1\ \mathrm{G\Omega}$ bias resistor to apply an AC bias current of $5\ \mathrm{nA}$ RMS at a frequency of $3.3373\ \mathrm{Hz}$. Keithley 2400 SourceMeters were used to apply voltages to the gates. All standard Hall configuration measurements were performed using the same voltage probes. One voltage contact behaved inconsistently and was not used in any of the measurements.

Hysteresis loops with respect to DC current bias were performed in zero magnetic field, unless otherwise noted. The DC current was applied using a Yokogawa 7651 DC voltage source with a $100\ \mathrm{M\Omega}$ bias resistor. The differential resistance was then determined by measuring the response to an additional $5\ \mathrm{nA}$ AC current added to the DC current bias. The AC current was sourced from the lock-in amplifier using a $1\ \mathrm{G\Omega}$ bias resistor as before, and both the AC and DC current sources were connected to the source terminal on the sample using a tee. Since the resistance from the source terminal to ground through the device was much smaller than either of the two bias resistors, the currents effectively added together.

\subsection*{Optical micrograph of the completed heterostructure}

\begin{figure}
    \centering
    \includegraphics[width=12cm]{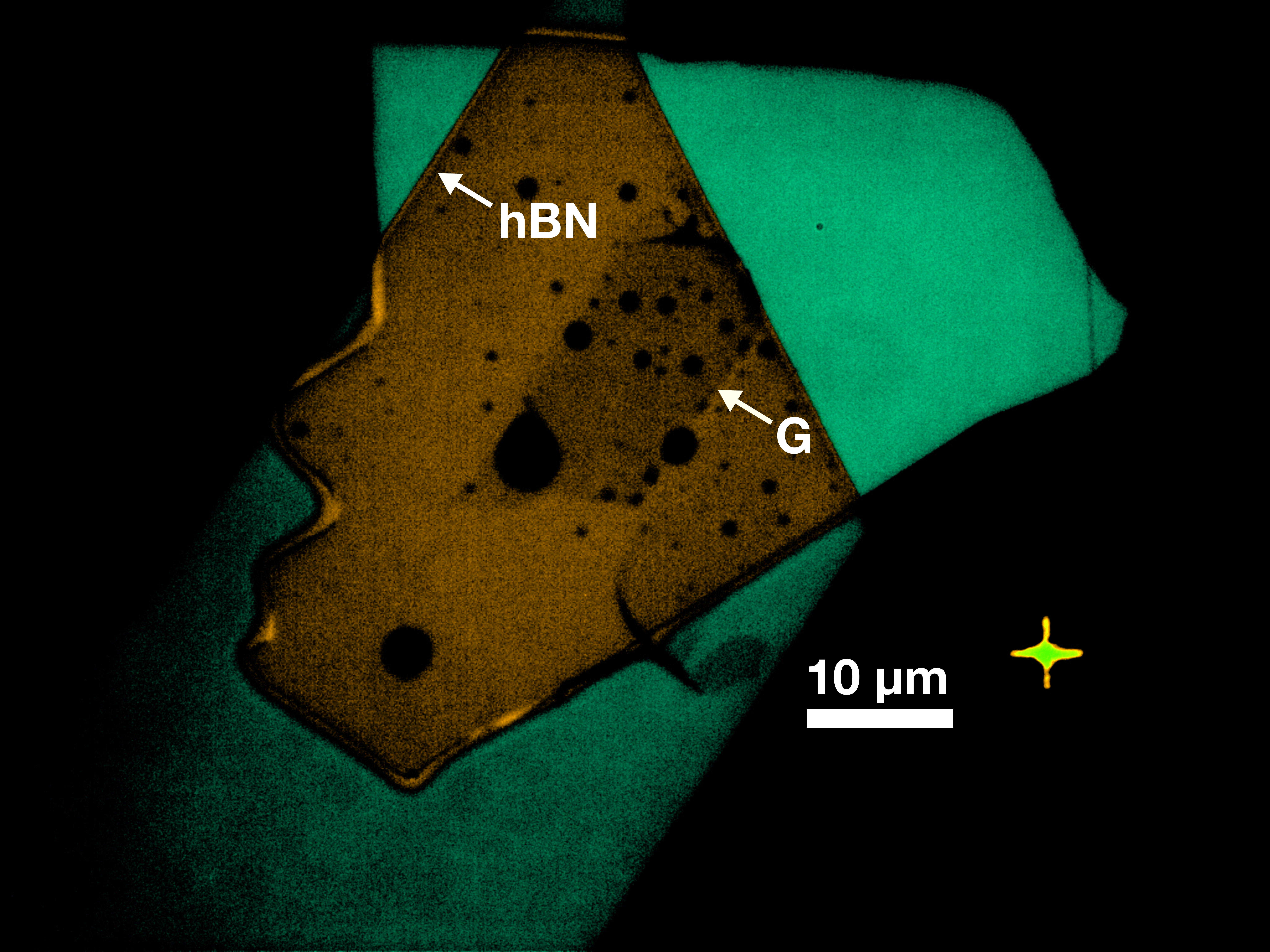}
    \caption{\textbf{Micrograph of the TBG heterostructure.}
    An optical micrograph of the completed heterostructure (before lithography) demonstrates the rotational alignment of the top hBN layer and graphene. The arrow labeled `G' indicates a crystallographic edge of the top graphene layer of the TBG while the arrow labeled `BN' indicates a crystallographic edge of the top encapsulating hBN crystal. Based on this micrograph, the alignment of the crystallographic edge of the top hBN to that of top graphene of the TBG is consistent with the experimentally measured angle of $\theta_{\mathrm{hBN}} = 0.83^\circ \pm 0.02^\circ$ (calculated based on the density corresponding to the peak we associate with the hBN moir\'e pattern; see Fig.~\ref{fig:fig1} of the main text). The bottom hBN is far from rotational alignment with the bottom graphene layer of the TBG.}
    \label{fig:micrograph}
\end{figure}

\subsection*{Variability of the displacement-field dependence between cooldowns}

\begin{figure}
    \centering
    \includegraphics[width=9.5cm]{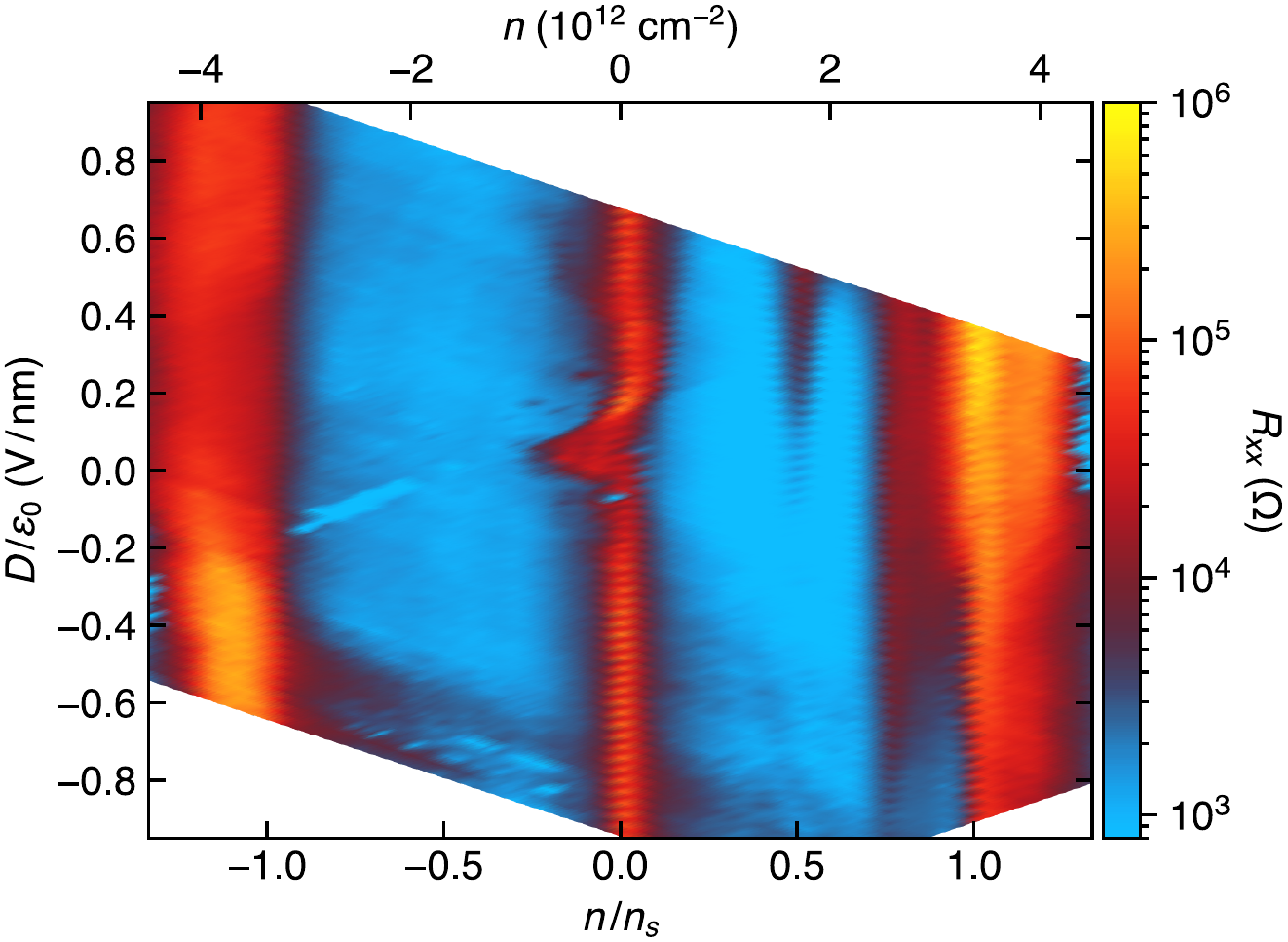}
    \caption{\textbf{Displacement field dependence from separate cooldown.} Longitudinal resistance $R_{xx}$ of the TBG device (measured between contacts separated by $2.15$ squares at $40\ \mathrm{mK}$) as a function of carrier density $n$ (shown on the top axis), filling factor relative to the superlattice density $n_s$ (bottom axis), and the applied perpendicular displacement field $D$ (left axis). This cooldown was separate from either of those described in the main text.}
    \label{fig:gatemap2}
\end{figure}

The dependence of the longitudinal resistance $R_{xx}$ on displacement field is not perfectly reproducible between cooldowns. Fig.~\ref{fig:gatemap2} shows a map of $R_{xx}$ as a function $n$ and $D$ at $T = 40\ \mathrm{mK}$ for a cooldown of the device separate from either of those described in the main text. When compared with Fig.~\ref{fig:fig1} of the main text, Fig~\ref{fig:gatemap2} appears to have an overall shift in $D$. In addition to this shift, there is an offset in the gate voltage corresponding to the resistance peak at the CNP that varies between cooldowns. We have accounted for this in calculating $n$ and $D$. Comparing to Fig.~\ref{fig:fig1} of the main text, this cooldown exhibits a broadening of the CNP resistance peak centered near $0.05~\mathrm{V/nm}$. Additionally, the position of the CNP resistance peak in the calculated density $n$ does not appear to exhibit a significant change with $D$, aside from asymmetric broadening near $D = 0$ that disappears at larger $|D|$ (cf. Fig.~\ref{fig:fig1} of the main text where the line traced by the peak in the 2D map appears to have a small kink). The differences between cooldowns are likely caused by variations in the disorder landscape seen by the TBG, which can be changed by thermal cycling the device. This variability between cooldowns did not impact our ability to observe magnetic hysteresis, which was present in each cooldown of the device in the dilution refrigerator.

\subsection*{Comparison with a second TBG device misaligned to hBN}

\begin{figure}
    \centering
    \includegraphics[width=15cm]{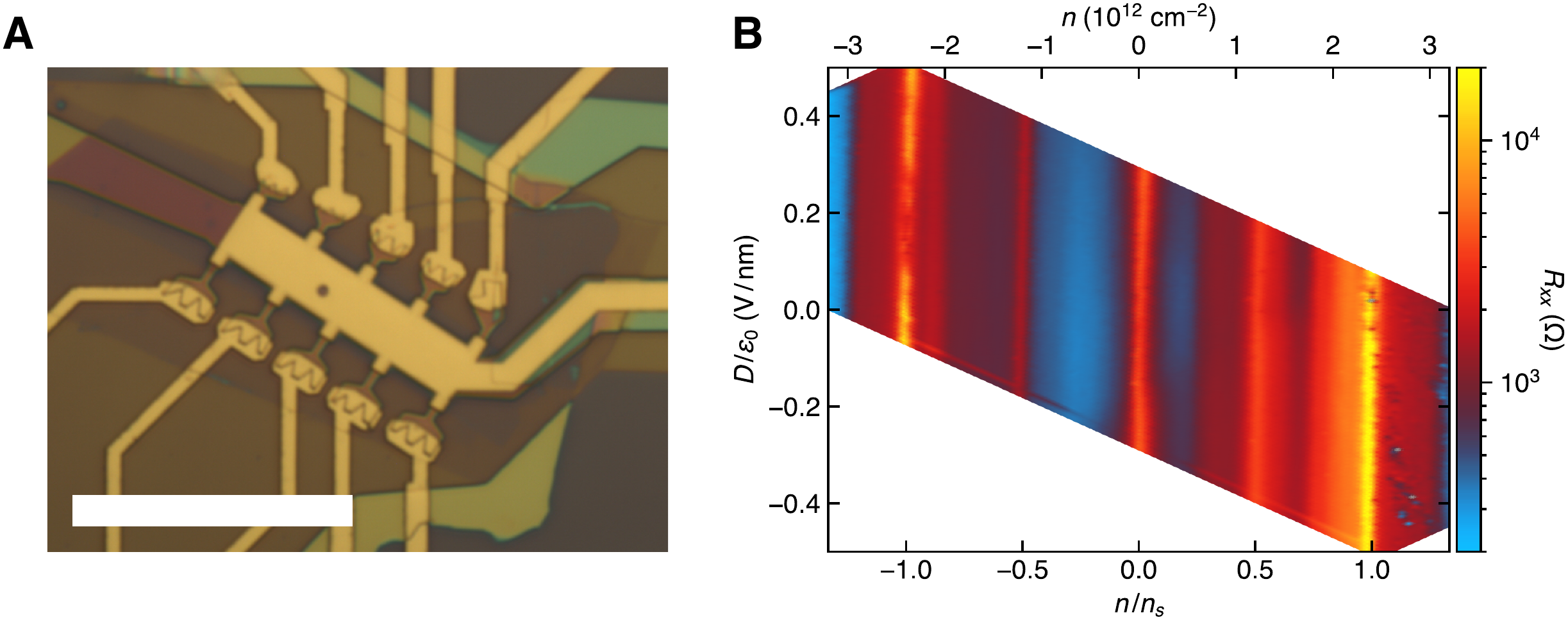}
    \caption{\textbf{Displacement field dependence of TBG device misaligned to hBN.}
    (\textbf{A}) An optical micrograph of the completed device. The scale bar is $20\ \mathrm{\mu m}$.
    (\textbf{B}) Longitudinal resistance $R_{xx}$ of the misaligned TBG device (measured between contacts separated by $1.25$ squares at $1.5\ \mathrm{K}$) as a function of carrier density $n$ (shown on the top axis, or as a filling factor relative to the superlattice density shown on the bottom axis) and perpendicular displacement field $D$.
    }
    \label{fig:2nd_dev}
\end{figure}

To explain the dependence of transport on the direction of the displacement field in the device presented in the main text, we can compare it to the dependence in a second TBG device (with a $1.05^\circ\pm 0.02^\circ$ twist angle) where the graphene has been intentionally misaligned with each of the hBN cladding layers. This second device has an additional graphite back gate which should help to drastically reduce disorder of the potential landscape within the TBG, by screening the effect of charges in the $\textrm{SiO}_\textrm{2}$~\cite{Hunt2013}. Also replacing the top metal gate with a second graphite gate could lead to further improvement~\cite{Zibrov2017, Yankowitz2018}. 

When the longitudinal resistance is measured as a function of the applied gate voltages, this device shows drastically different dependence on the applied displacement field than the device of the main text. With the graphene misaligned with both hBN layers, the device is in an approximately symmetric dielectric environment (up to differences in disorder, hBN thicknesses, and gate materials). As a result, the longitudinal resistance has no features that strongly depend on the direction of the displacement field.

The stark contrast between the displacement field dependence of this misaligned device and the device of the main text suggests that the vertical symmetry of the heterostructure is broken in the latter. This could result from disorder, but as discussed in the main text, it appears that the symmetry is broken by the alignment of one of the two hBN cladding layers with the TBG. The device with misaligned hBN also did not show AH effect at any filling.

\subsection*{Repeatability of hysteresis loops}

\begin{figure}
    \centering
    \includegraphics[width=9cm]{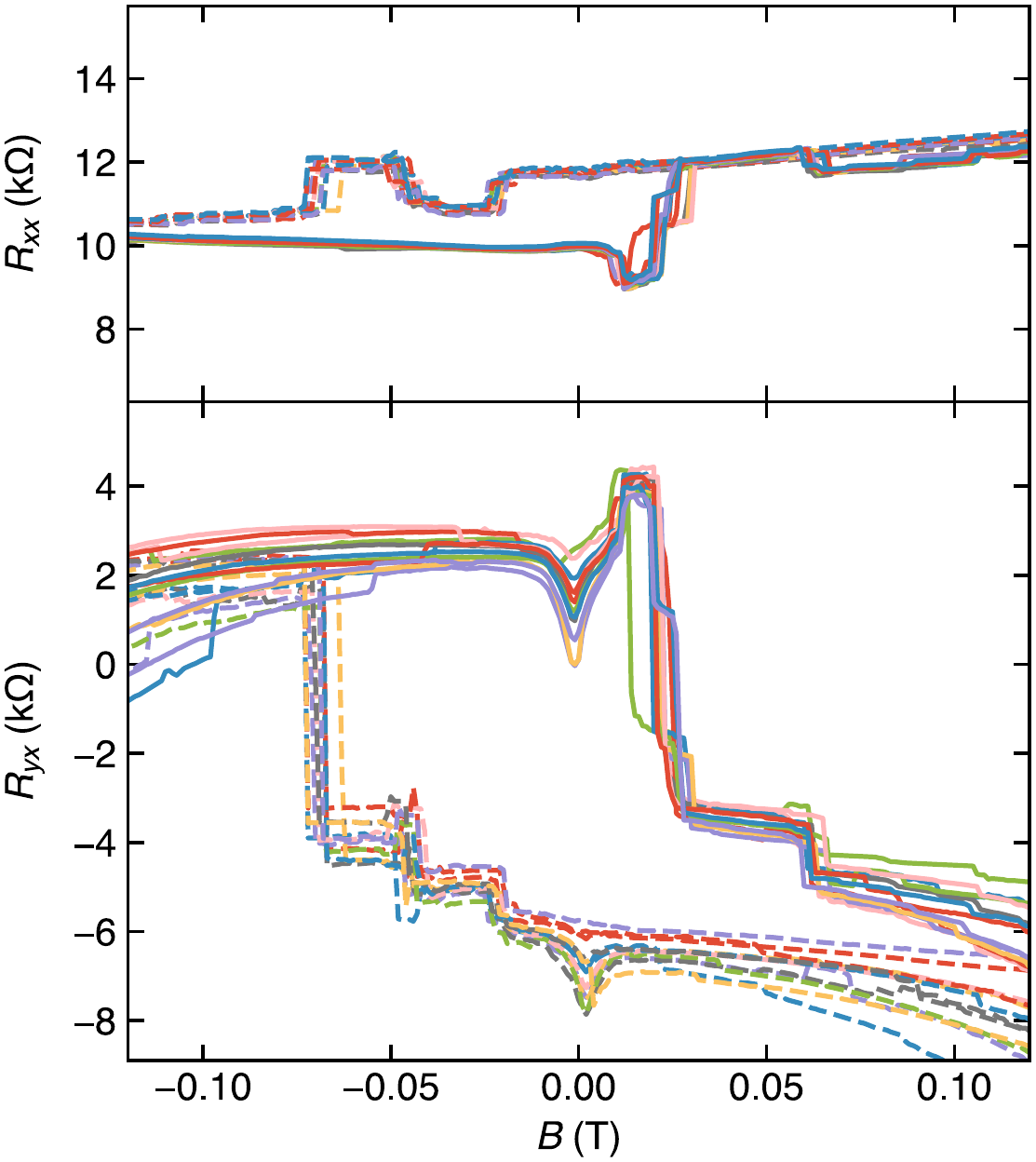}
    \caption{\textbf{Repeated hysteresis loops.}
    Longitudinal resistance $R_{xx}$ (top panel) and Hall resistance $R_{yx}$ (bottom panel) are shown as a function of magnetic field for twelve consecutive loops of the field between $\pm 250\ \mathrm{mT}$ for $n/n_s = 0.758$ and $D/\epsilon_0 = 0\ \mathrm{V/nm}$ (in the same cooldown as that of Fig.~\ref{fig:gatemap2}). The solid and dashed lines correspond to measurements taken while sweeping the magnetic field $B$ up and down, respectively.}
    \label{fig:hyst_repeat}
\end{figure}
To check the repeatability of the magnetic field dependence of transport, we sweep the applied magnetic field between $\pm 250\ \mathrm{mT}$ twelve times while maintaining a constant density and displacement field (Fig.~\ref{fig:hyst_repeat}). The structure of the hysteresis loop is very consistent between sweeps with many of the intermediate jumps appearing highly repeatable. These jumps likely correspond to some magnetic domain structure with domains of varying coercivities, or to a repeatable pattern of domain wall motion and pinning. As discussed in the main text, this behavior may result from inhomogeneity caused by local variations in effective gating or in the twist angle between the graphene sheets.
Additionally, we see a dip (of unknown origin) in the Hall resistance on the advanced side of zero field for both directions of the hysteresis loop.

\subsection*{Extracting anomalous and conventional components of Hall effect}

To clearly separate the two components of the Hall effect, we define the anomalous component as the difference between $B=0$ Hall signals on the up and down sweeps of applied field:
$R_{yx}^\text{AH}= \left|R_{yx}^\uparrow - R_{yx}^\downarrow \right|/2$, where $R_{yx}^{\uparrow (\downarrow)}$ is the Hall resistance remaining at zero field after the sample has been magnetized by an upward (downward) applied field. If we took $R_{yx}^{\uparrow (\downarrow)}$ to equal the raw values of $R_{yx}(B=0)$ under the two sweep directions, our reported parameters would be affected by small jumps in the resistance close to zero field. Instead, in analyzing the data, we examine a sweep toward zero field from a positive (negative) field higher than the coercive field; starting at a chosen high field cutoff, we include data until the first significant jump in $R_{yx}$. We fit a line to this subset and extrapolate or interpolate the value at $B=0$ to estimate $R_{yx}^{\uparrow (\downarrow)}$. By varying the high field cutoff, we can estimate the uncertainty in the fitting parameters. We use the slope of these same linear fits as a measure of the conventional component of the Hall signal, $R_H$.

\clearpage
\subsection*{Extended discussion on the nature of the observed AH effect}

In this section, we further develop our discussion of the source of the AH effect we have observed, and explore why it is not quantized. As mentioned in the main text, an AH effect can either be intrinsic, arising from Berry curvature of the filled bands, or extrinsic, resulting from scattering mechanisms.

\begin{figure}
    \centering
    \includegraphics[width=11cm]{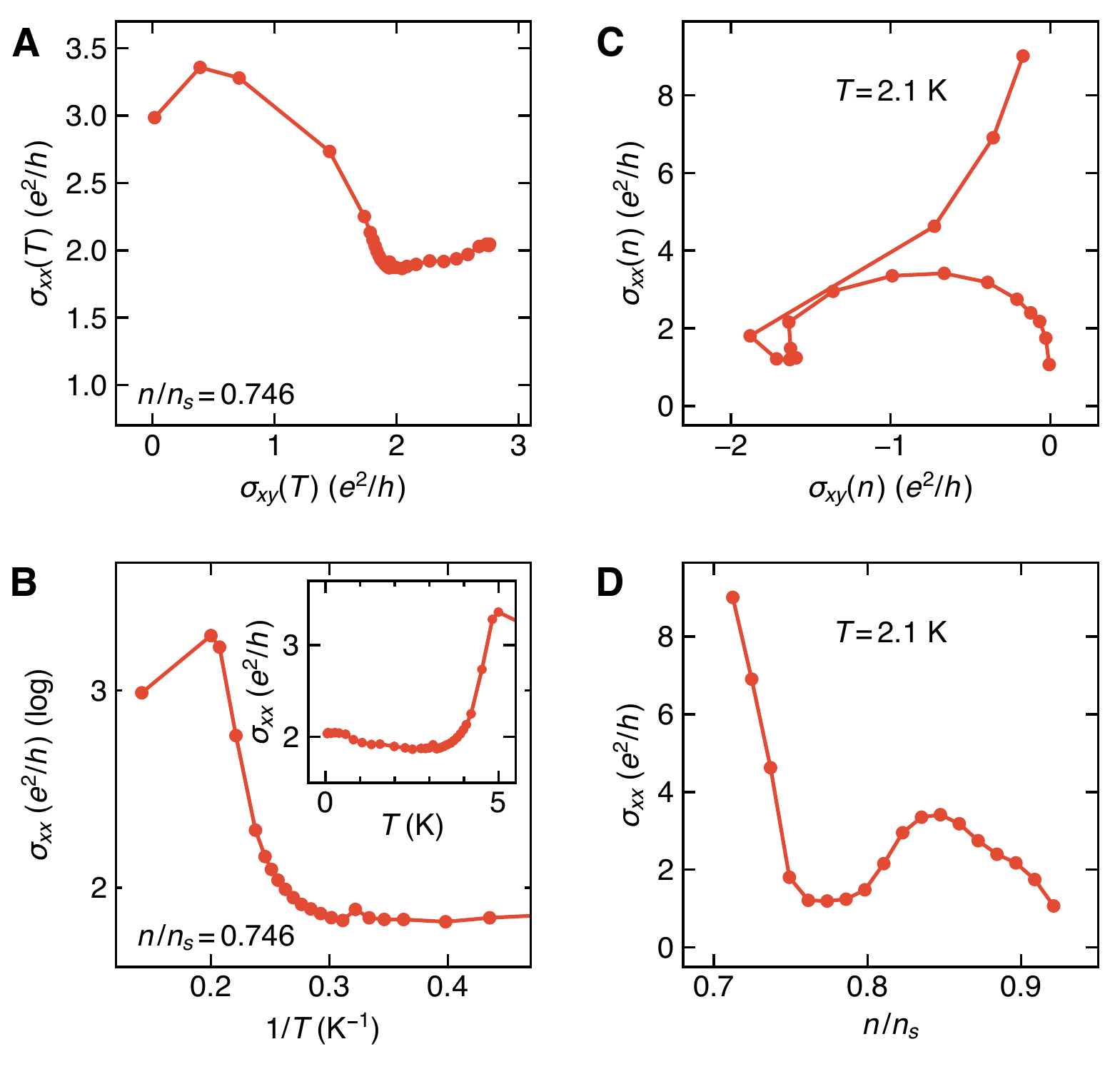}
    \caption{\textbf{Behavior of the conductivity tensor.}
    (\textbf{A}) The longitudinal conductivity $\sigma_{xx}$ is plotted parametrically against the Hall conductivity $\sigma_{yx}$ for a series of measurements at different temperatures with the density fixed at $n/n_s = 0.746$ and $D/\epsilon_0 = -0.62\ \mathrm{V/nm}$ (shown in Fig.~\ref{fig:fig2}C of the main text). All conductivity values in this figure have been extracted from resistance measurements taken at $50\ \mathrm{mT}$ when sweeping the applied field downward from a value larger than the coercive field (so the sample has been magnetized by an upward field). The resistivity is derived from the measurements by assuming a homogeneous sample, and the conductivities are given by $\sigma_{xx} = \rho_{xx}/\left(\rho_{xx}^2+\rho_{yx}^2\right)$ and $\sigma_{xy} = \rho_{yx}/\left(\rho_{xx}^2+\rho_{yx}^2\right)$. The relationship between $\sigma_{xy}$ and $\sigma_{xx}$ is not consistent with an extrinsic AH effect resulting from skew scattering.
    (\textbf{B}). Arrhenius plot of $\sigma_{xx}$ on a log scale versus $1/T$, with the same data shown in (A).
    (Inset) $\sigma_{xx}$ is plotted on a linear scale against temperature.
    (\textbf{C}) $\sigma_{xx}$ is plotted parametrically against $\sigma_{xy}$ for a series of measurements at different densities at $T = 2.1\ \mathrm{K}$, with $D/\epsilon_0 = -0.22\ \mathrm{V/nm}$. These data were obtained during the same cooldown as that of the data shown in Figs.~\ref{fig:fig1}, \ref{fig:fig3}, and \ref{fig:fig4} of the main text. Again, the behavior appears inconsistent with skew scattering. Moreover, it is qualitatively similar to the density dependence of the conductivity in a magnetic topological insulator approaching a QAH effect shown in Ref.~\citen{Checkelsky2014}.
    (\textbf{D}) $\sigma_{xx}$ as a function of temperature, from the same data as in (C), showing the emergence of a dip in $\sigma_{xx}$ around $n/n_s = 3/4$ consistent with the approach to a Chern insulator state.}
    \label{fig:sigma}
\end{figure}

One cause of an extrinsic AH effect, skew scattering, is associated with a linear relationship $\sigma_{xy} \propto \sigma_{xx}$ (where $\sigma_{xy}$ and $\sigma_{xx}$ are the Hall and longitudinal conductivities, respectively), which is clearly inconsistent with our data: for datasets parameterized by either $n/n_s$ or temperature, shown in Fig.~\ref{fig:sigma} A and C, this relationship is highly nonlinear. However, distinguishing between intrinsic Berry curvature and the other extrinsic mechanism, side jump scattering, is more challenging. Generally, the observed AH effect is compared with the theoretical expectation for the intrinsic contribution~\cite{Nagaosa2010}, but a clear theoretical consensus does not yet exist in the literature in this case.

Instead, we have argued in the main text that the size of $\rho_{yx}^\text{AH}$ compared to that in other AH materials is evidence that its source is intrinsic. We reiterate and expand on this comparison here. As mentioned, the largest Hall angle measured in our device is $\rho_{xy}/\rho_{xx}=1.4$, whereas previously reported (extrinsic or intrinsic) AH materials yield $\rho_{xy}/\rho_{xx}\lesssim 0.2$~\cite{Liu2018}, except for the magnetic topological insulators exhibiting near-vanishing $\rho_{xx}$ in the QAH effect. Further, we found $\rho_{yx}^\text{AH}$ as large as $0.4h/e^2$, greater than in early magnetic topological insulators~\cite{Checkelsky2012,Chang2013a,Kou2013}. The corresponding longitudinal resistivity of $0.3h/e^2$ is also comparable to that of the first samples of those same materials to display near-quantization of $\rho_{yx}$ in zero field $(\rho_{yx}\ge 0.98 h/e^2)$~\cite{Chang2013b,Checkelsky2014}.

We have also argued in the main text that the origin is topological based on the appearance of the AH effect at an apparent insulating state. When the Fermi level is in a gap at zero temperature, the extrinsic mechanisms cannot contribute to $\sigma_{xy}$~\cite{Nagaosa2010}. In these conditions, the system is in a Chern insulating state if the occupied bands carry a net Chern number, or in a trivial insulating state with $\sigma_{xy}=0$ otherwise. As we have discussed, the state at three-quarters filling is evidently not a single-domain, ideal Chern insulator given that $\rho_{xx}$ does not vanish. However, we believe that our data could be consistent with the presence of a small, topologically nontrivial gap. One piece of evidence for this is that the ordinary Hall slope passes through zero and changes sign near the value of $n$ at which $\rho_{yx}^\text{AH}$ is maximized (see Fig.~\ref{fig:fig2}B of the main text). A corresponding dip is seen in $\sigma_{xx}$ as a function of $n/n_s$, shown in Fig.~\ref{fig:sigma}D.

In other studies of near-magic-angle TBG samples, a high-resistance state has also been observed at $n/n_s = 3/4$~\cite{Yankowitz2018,Cao2018b}. They did not report large AH effect. The difference may be due to the alignment of the TBG with hBN in our experiment, as mentioned in the main text.

With nonlocal measurements also suggestive of edge state transport, we believe our collection of evidence suggests the system is approaching a Chern insulating state, but it remains to explain why it is non-ideal. There could be several reasons $\rho_{yx}$ is not quantized and $\rho_{xx}$ is nonzero, contrary to the expectation for a QAH effect. For one, the 2D bulk may not be strongly insulating, leading to both a nonzero $\rho_{xx}$ and a reduction in the measured value of $\rho_{yx}$~\cite{Chang2013a,Checkelsky2012}.

Exploring the possibility of bulk conduction via the temperature dependence of $\sigma_{xx}$, we find that below $3\ \mathrm{K}$ some conduction mechanism persists, insensitive to temperature (Fig.~\ref{fig:sigma}B). Between $3\ \mathrm{K}$ and $5\ \mathrm{K}$, conductivity rises with increasing temperature, but this is not a large enough range to allow us to identify possible activated behavior in parallel with the temperature-insensitive conduction. Above $5\ \mathrm{K}$ conductance stops rising. The low temperature residual conduction may be explained by inhomogeneity, which may result from local variations in twist angle or density, as we have noted elsewhere. Because of spatial variations in the density or gap size, it may be impossible to achieve a state in this sample with the Fermi level uniformly in the gap. Additionally, if such inhomogeneity results in a mixed domain state, edge states at domain walls could form a complex network resulting in nonzero $\rho_{xx}$ and non-quantized $\rho_{yx}$ even at zero temperature, and even if the magnitude of the gap were uniform (see Refs.~\citen{Rosen2017} and~\citen{Yasuda2017} for examples of very simple mixed domain configurations yielding significant $\rho_{xx}$). Given the signs of inhomogeneity we have observed, we suspect at least one of these scenarios may be realized in our sample. Even though such effects prevent the observation of a clear QAH effect, our results laid out in the main text and above nonetheless point to the existence of a topologically nontrivial gap opened by interactions.

To be clear: the experimental data we present do not unambiguously demonstrate that the three-quarters state is a Chern insulator. A non-quantized AH effect could result from Berry curvature even if the Fermi level is not in a gap, though, as discussed above, the magnitude of the Hall angle observed here far exceeds that of any reported AH system not known to host a Chern insulator state.

Another possibility is an AH analogue to the Hall insulator~\cite{Kivelson1992,Hilke1998}. In such a state, as $T \to 0$, $\rho_{xx} \to \infty$ while $\rho_{xy}$ remains finite and nonzero, corresponding to $\sigma_{yx} \propto \sigma_{xx}^2 \to 0$. This alternative might be plausible given that the peak in $R_{yx}^\text{AH}$ (Fig.~\ref{fig:fig2} of the main text) occurs at a peak in $R_{xx}$ (Fig.~\ref{fig:fig1} of the main text) rather than at a minimum of $R_{xx}$ as would be expected for a Chern insulating state. However, when we attempt to extract conductivities from our resistivity measurements, their temperature dependence appears inconsistent with a Hall insulator: with decreasing temperature, $\sigma_{xx}$ initially falls but then saturates (Fig.~\ref{fig:sigma}B), and $\sigma_{yx}$ increases (Fig.~\ref{fig:sigma}A). Furthermore, the Hall insulator state is not known to host topologically protected edge states, the presence of which is suggested by our nonlocal transport measurements.


\subsection*{Stability of the magnetization}

\begin{figure}
    \centering
    \includegraphics[width=15cm]{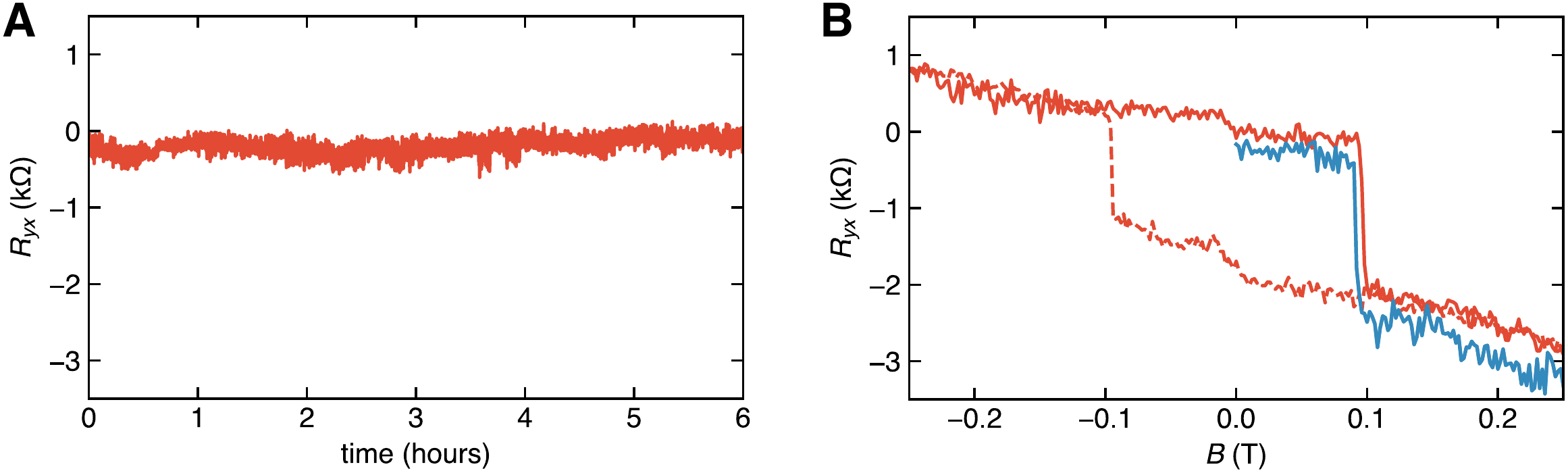}
    \caption{
    \textbf{Temporal stability of the magnetization}
    (\textbf{A}) 
    Hall resistance $R_{yx}$ at $n/n_s = 0.746$ and $D/\epsilon_0 = -0.62\ \mathrm{V/nm}$ as a function of time over the course of $6$ hours in zero field, after first magnetizing the sample by applying $-250~\mathrm{mT}$ and then returning the field to $0~\mathrm{T}$.
    (\textbf{B})
    A full hysteresis loop taken prior to the measurement shown in (A) is displayed in red. The blue trace shows the behavior of $R_{yx}$ as the field is swept from $0$ to $250~\mathrm{mT}$ following the measurement in (A). A clear anomalous Hall jump in the blue trace is comparable to those in the continuous red loop, indicating that the magnetization was stable through the $6$~hour pause.}
    \label{fig:stability}
\end{figure}

\subsection*{Additional density dependence of hysteresis loops and effect of displacement field}

\begin{figure}
    \centering
    \includegraphics[width=15cm]{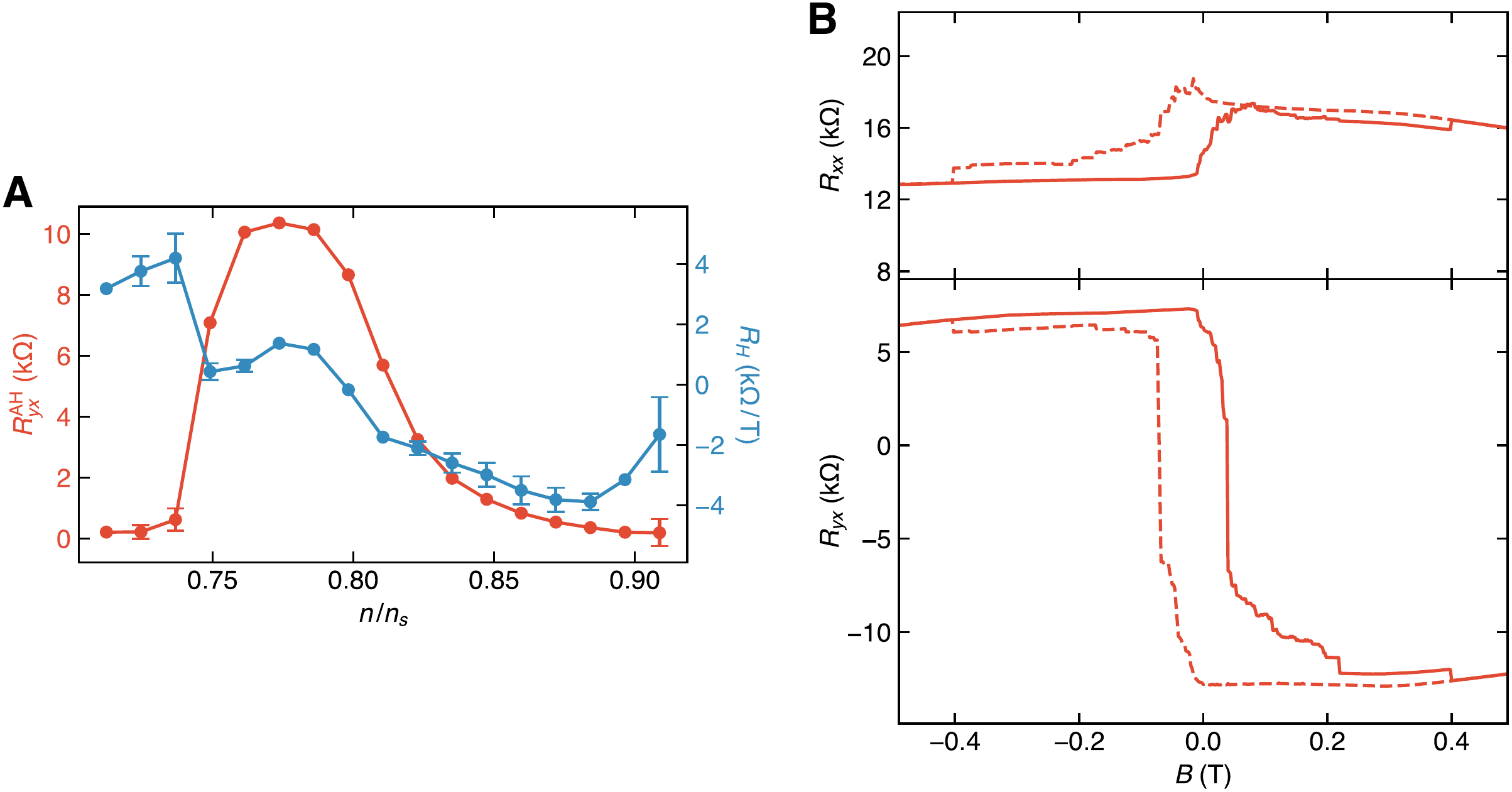}
    \caption{
    \textbf{Density dependence near $3/4$ with fixed displacement field at $2.1\ \mathrm{K}$} (\textbf{A}) 
    Zero-field anomalous Hall resistance $R_{yx}^\text{AH}$ (red) and ordinary Hall slope $R_H$ (blue) as a function of $n/n_s$ while maintaining a constant displacement field $D/\epsilon_0=-0.22\ \mathrm{V/nm}$. $R_{yx}^\text{AH}$ is peaked at $n/n_s = 0.774$, close to the 0.76 in Fig.~\ref{fig:fig2}B of the main text and again coincident with a sign change in $R_H$. The full width at half maximum is slightly increased, at $0.07$ instead of $0.04$.
    (\textbf{B})~Magnetic field dependence of the longitudinal resistance $R_{xx}$ (upper panel) and Hall resistance $R_{yx}$ (lower panel) at $n/n_s=0.774$, the largest hysteresis loop of the series shown in (A), with $R_{yx}^\text{AH}=10.4\ \mathrm{k\Omega}$.}
    \label{fig:2k_loops}
\end{figure}

\begin{figure}
    \centering
    \includegraphics[width=10cm]{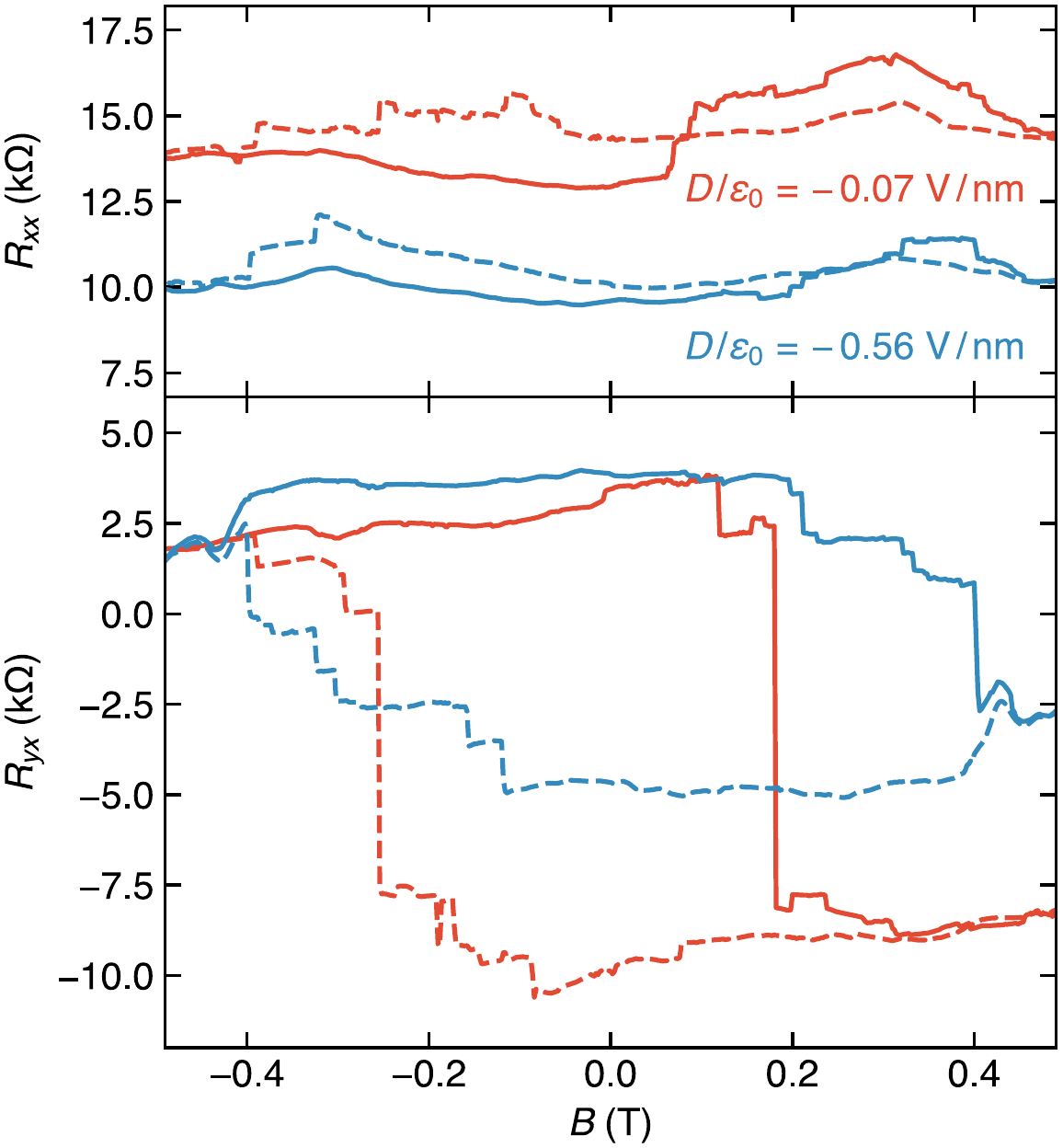}
    \caption{
    \textbf{Displacement field dependence of hysteresis loops}
    Longitudinal resistance $R_{xx}$ (upper panel) and Hall resistance $R_{yx}$ (lower panel) at $n/n_s=0.749$ for two different displacement fields as labeled in the figure. Although tuning the displacement field from a large negative field to near zero causes a slight change in the longitudinal resistance and the hysteresis loop structure, the TBG magnetic field dependence remains hysteretic.}
    \label{fig:loop_d_dep}
\end{figure}

\subsection*{Additional characterization of current-driven switching}

\begin{figure}
    \centering
    \includegraphics[width=14cm]{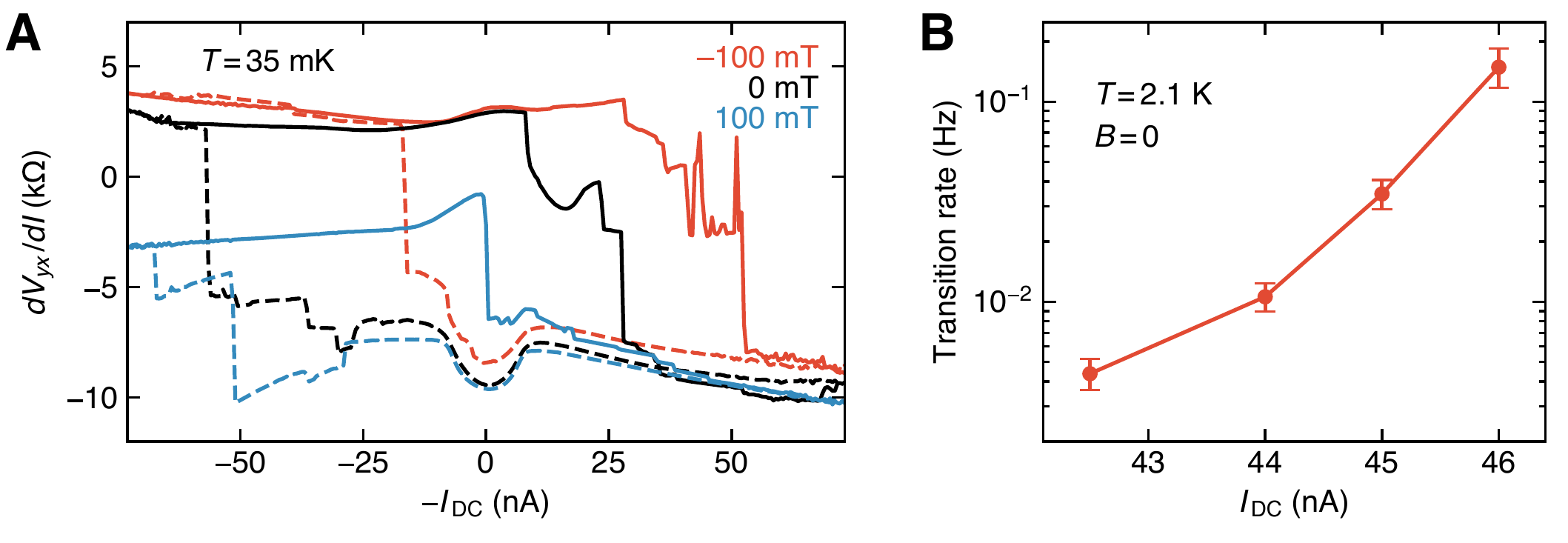}
    \caption{\textbf{Current-driven switching in nonzero magnetic field, and characterization of the transition.}
    (\textbf{A}) Hysteresis loops of the differential Hall resistance $dV_{yx}/dI$ with respect to DC current (plotted as $-I_\text{DC}$ as in Fig.~\ref{fig:fig4} of the main text) at three different static magnetic fields after the sample was magnetized at $500\ \mathrm{mT}$. These data were taken at $35\ \mathrm{mK}$ with $n/n_s = 0.749$ and $D/\epsilon_0 = -0.22\ \mathrm{V/nm}$ during the same cooldown as for the data of Fig.~\ref{fig:fig4}.
    (\textbf{B}) Transition rate of the apparent magnetization switching at a fixed current $I_{DC}$ after magnetizing the sample with a $-75\ \mathrm{nA}$ current (at $T = 2.1\ \mathrm{K}$ and zero field). The transition appears to be a memoryless process.
    }
    \label{fig:dcloops}
\end{figure}

The dependence of the DC current hysteresis loop (shown at zero field at $T = 2.1\ \mathrm{K}$ in Fig.~\ref{fig:fig4} of the main text) on applied field at $T = 35\ \mathrm{mK}$ is shown in Fig.~\ref{fig:dcloops}A. For each loop, the magnetic field was first increased to $500\ \mathrm{mT}$ and then decreased to the target field before cycling the DC current $I_\text{DC}$. Evidently, the applied field shifts the critical current required to switch the differential Hall resistance $dV_{yx}/dI$. Consistent with our expectation, when a nonzero field is applied, the magnitude of the current required to switch the apparent magnetization to the direction opposite (aligned) with the field is increased (decreased). It also appears from the loop at $100\ \mathrm{mT}$ that with sufficient field in the direction in which the sample has been magnetized, the DC current cannot completely reverse the magnetization, since $dV_{yx}/dI$ remains significantly lower after switching than it does in the loops at $0$ or $-100\ \mathrm{mT}$.

We further attempted to study the dynamics of the switching transition by measuring the time dependence of the differential Hall signal when $I_\text{DC}$ was close to a value at which we observe large jumps in the loop shown in Fig.~\ref{fig:fig4} of the main text. With the current fixed near the jump at ${\sim}45\ \mathrm{nA}$, $dV_{yx}/dI$ appears stable for a short time before rapidly switching (we did not measure the switching time itself since our lock-in measurements could not resolve changes on a time scale faster than ${\sim}1\ \mathrm{s}$).
By repeatedly bringing $I_\text{DC}$ to $-75\ \mathrm{nA}$ and back to a target current near the jump (at a rate of $0.7\ \mathrm{nA/s}$), then measuring $dV_{yx}/dI$ as a function of time, we were able to investigate the statistics of this time lapse before the transition occurred. We find that the delay time before the switch appears to be exponentially distributed, indicating that the transition is a memoryless process that does not depend on the total charge transported. The corresponding transition rate (Fig.~\ref{fig:dcloops}B) rapidly increases with $I_{DC}$ near the currents at which we observe jumps in the full hysteresis loops, which were obtained by sweeping $I_{DC}$ slowly at an average rate of $0.15\ \mathrm{nA/s}$.

\subsection*{Quantum oscillations}

\begin{figure}
    \centering
    \includegraphics[width=\textwidth]{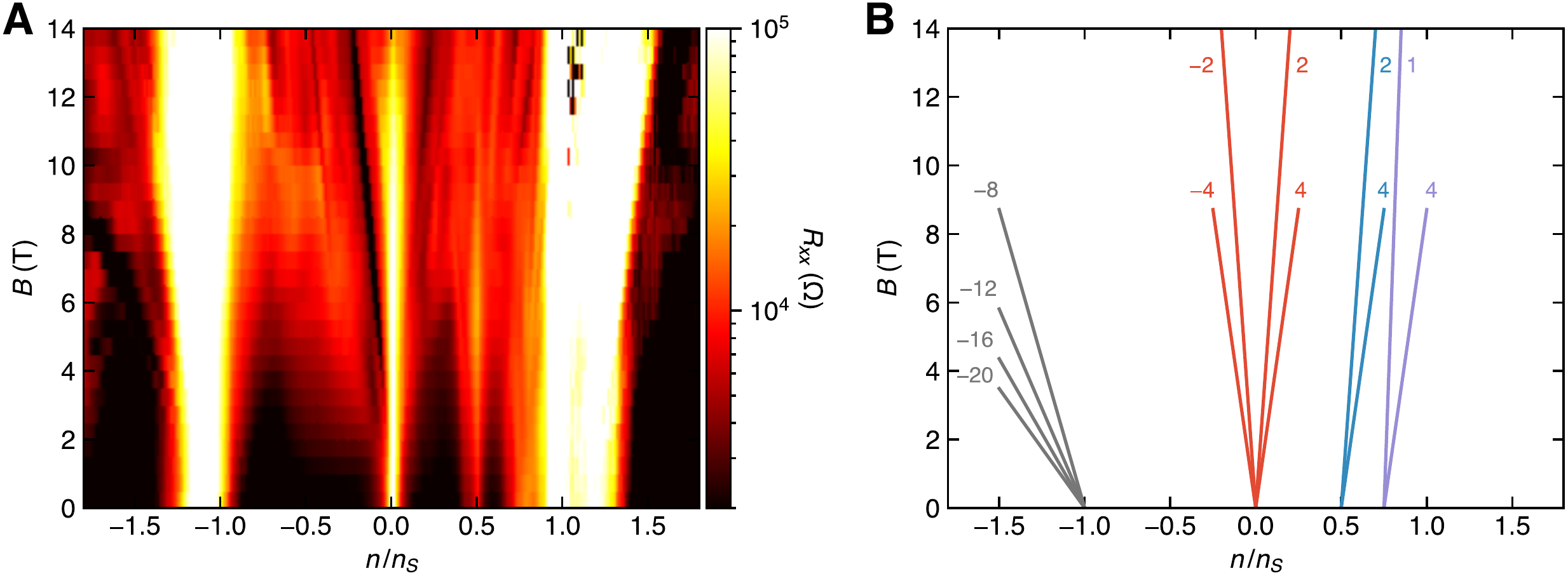}
    \caption{\textbf{Quantum oscillations of TBG at fixed displacement field} 
    (\textbf{A}) 
    Landau fan diagram of the longitudinal resistance $R_{xx}$ taken at $2.1\ \mathrm{K}$ for a fixed displacement field $D\epsilon_0 = 0\ \mathrm{V/nm}$. Emerging from the CNP, we observe the landau levels $\nu = \pm 2, \pm4$. We further observe landau levels from $n/n_s = 1/2$ of $\nu = 2,4$, from $n/n_s=3/4$ of $\nu=1,4$, and the sequence from $n/n_s=-1$ of $\nu = -8,-12,-16,...$.  
    (\textbf{B}) 
    Schematic of the Landau levels observed in (A).}
    \label{fig:fan}
\end{figure}

We observe several sets of Landau fans emerging from the high resistance states of the TBG (Fig.~\ref{fig:fan}A). The discernible quantum oscillations are represented schematically in Fig.~\ref{fig:fan}B. The degeneracy of Landau levels is representative of the symmetries of the electronic band structure and may yield information about where spin, valley, or layer symmetry may be broken.

Emerging from the CNP, the $\nu = \pm2$ and $\pm4$ Landau levels are clearly visible. This is slightly surprising as one might expect the level sequence to reflect spin and valley degeneracies, and hence four-fold degenerate Landau levels, as has been seen previously in magic angle TBG samples~\cite{Cao2018b,Yankowitz2018}. In the fan emerging from $n/n_s = 0.5$, we see similar periodicity with signs of $\nu = 2$ and $4$. Additionally, there is a strong enhancement of the resistance at $n/n_s=0.5$ with the applied out-of-plane magnetic field, peaking at roughly 6~T, in contrast to the behavior of the apparent Mott insulating state observed by Cao \emph{et al.}~\cite{Cao2018a}. The $n/n_s = 0.75$ fan exhibits a further reduction of the periodicity such that we only observe clear signatures of $\nu = 1$ and $4$. This reduction in symmetry at $n/n_s=0.75$ could reflect interaction-driven lifting of the degeneracies: if the conduction bands for three spin-valley flavors are fully filled, we may be observing the quantum oscillations of a single flavor. No clear quantum oscillations emerge from $n/n_s=1$, but the presence of $\nu = -8,-12,$, $-16$, and $-20$ levels originating from $n/n_s = -1$ is consistent with what's seen in other TBG samples~\cite{Cao2018b,Yankowitz2018}.


\begin{thebibliography}{10}
	
	\bibitem{Bistritzer2011}
	R.~Bistritzer, A.~H. Macdonald, {\it Proc. Natl. Acad. Sci. U.S.A.\/} {\bf
		108}, 12233 (2011).
	
	\bibitem{Fang2016}
	S.~Fang, E.~Kaxiras, {\it Phys. Rev. B\/} {\bf 93}, 235153 (2016).
	
	\bibitem{Nam2017}
	N.~N. Nam, M.~Koshino, {\it Phys. Rev. B\/} {\bf 96}, 075311 (2017).
	
	\bibitem{Cao2016}
	Y.~Cao, {\it et~al.\/}, {\it Phys. Rev. Lett.\/} {\bf 117}, 116804 (2016).
	
	\bibitem{Cao2018a}
	Y.~Cao, {\it et~al.\/}, {\it Nature\/} {\bf 556}, 80 (2018).
	
	\bibitem{Yankowitz2018}
	M.~Yankowitz, {\it et~al.\/}, {\it
		\normalfont{https://arxiv.org/abs/1808.07865}\/}  (2018).
	
	\bibitem{Cao2018b}
	Y.~Cao, {\it et~al.\/}, {\it Nature\/} {\bf 556}, 43 (2018).
	
	\bibitem{Xie2018}
	M.~{Xie}, A.~H. {MacDonald}, {\it
		\normalfont{https://arxiv.org/abs/1812.04213}\/}  (2018).
	
	\bibitem{Ochi2018}
	M.~Ochi, M.~Koshino, K.~Kuroki, {\it Phys. Rev. B\/} {\bf 98}, 081102 (2018).
	
	\bibitem{Dodaro2018}
	J.~F. Dodaro, S.~A. Kivelson, Y.~Schattner, X.~Q. Sun, C.~Wang, {\it Phys. Rev.
		B\/} {\bf 98}, 075154 (2018).
	
	\bibitem{Thomson2018}
	A.~Thomson, S.~Chatterjee, S.~Sachdev, M.~S. Scheurer, {\it Phys. Rev. B\/}
	{\bf 98}, 075109 (2018).
	
	\bibitem{Kim2016}
	K.~Kim, {\it et~al.\/}, {\it Nano Lett.\/} {\bf 16}, 1989 (2016).
	
	\bibitem{Sharpe2018Supplement}
	{\bf \normalfont{See supplementary materials}}.
	
	\bibitem{Oostinga2008}
	J.~B. Oostinga, H.~B. Heersche, X.~Liu, A.~F. Morpurgo, L.~M.~K. Vandersypen,
	{\it Nat. Mater.\/} {\bf 7}, 151 (2008).
	
	\bibitem{Hunt2013}
	B.~Hunt, {\it et~al.\/}, {\it Science\/} {\bf 340}, 1427 (2013).
	
	\bibitem{Yoo2018}
	H.~Yoo, {\it et~al.\/}, {\it \normalfont{https://arxiv.org/abs/1804.03806}\/}
	(2018).
	
	\bibitem{Xue2011}
	J.~Xue, {\it et~al.\/}, {\it Nat. Mater.\/} {\bf 10}, 282 (2011).
	
	\bibitem{Emori2015}
	S.~Emori, C.~K. Umachi, D.~C. Bono, G.~S. Beach, {\it J. Magn. Magn. Mater.\/}
	{\bf 378}, 98 (2015).
	
	\bibitem{Nagaosa2010}
	N.~Nagaosa, J.~Sinova, S.~Onoda, A.~H. MacDonald, N.~P. Ong, {\it Rev. Mod.
		Phys.\/} {\bf 82}, 1539 (2010).
	
	\bibitem{Chang2013b}
	C.-Z. Chang, {\it et~al.\/}, {\it Science\/} {\bf 340}, 167 (2013).
	
	\bibitem{Checkelsky2014}
	J.~G. Checkelsky, {\it et~al.\/}, {\it Nat. Phys.\/} {\bf 10}, 731 (2014).
	
	\bibitem{Kou2014}
	X.~Kou, {\it et~al.\/}, {\it Phys. Rev. Lett.\/} {\bf 113}, 137201 (2014).
	
	\bibitem{Fox2018}
	E.~J. Fox, {\it et~al.\/}, {\it Phys. Rev. B\/} {\bf 98}, 075145 (2018).
	
	\bibitem{Yu2010}
	R.~Yu, {\it et~al.\/}, {\it Science\/} {\bf 329}, 61 (2010).
	
	\bibitem{Bestwick2015}
	A.~J. Bestwick, {\it et~al.\/}, {\it Phys. Rev. Lett.\/} {\bf 114}, 187201
	(2015).
	
	\bibitem{Chang2015}
	C.-Z. Chang, {\it et~al.\/}, {\it Phys. Rev. Lett.\/} {\bf 115}, 057206 (2015).
	
	\bibitem{Buttiker1988}
	M.~B{\"{u}}ttiker, {\it Phys. Rev. B\/} {\bf 38}, 9375 (1988).
	
	\bibitem{vanderPauw1958}
	L.~J. van~der Pauw, {\it Philips Res. Rep.\/} {\bf 13}, 1 (1958).
	
	\bibitem{Wang2013a}
	J.~Wang, B.~Lian, H.~Zhang, S.-C. Zhang, {\it Phys. Rev. Lett.\/} {\bf 111},
	086803 (2013).
	
	\bibitem{Rosen2017}
	I.~T. Rosen, {\it et~al.\/}, {\it npj Quantum Mater.\/} {\bf 2}, 69 (2017).
	
	\bibitem{Yasuda2017}
	K.~Yasuda, {\it et~al.\/}, {\it Science\/} {\bf 358}, 1311 (2017).
	
	\bibitem{Apalkov2016}
	D.~Apalkov, B.~Dieny, J.~M. Slaughter, {\it Proc. IEEE\/} {\bf 104}, 1796
	(2016).
	
	\bibitem{Upadhyaya2016}
	P.~Upadhyaya, Y.~Tserkovnyak, {\it Phys. Rev. B\/} {\bf 94}, 020411 (2016).
	
	\bibitem{Liu2018}
	E.~Liu, {\it et~al.\/}, {\it Nat. Phys.\/} {\bf 14}, 1125 (2018).
	
	\bibitem{Checkelsky2012}
	J.~G. Checkelsky, J.~Ye, Y.~Onose, Y.~Iwasa, Y.~Tokura, {\it Nat. Phys.\/} {\bf
		8}, 729 (2012).
	
	\bibitem{Chang2013a}
	C.-Z. Chang, {\it et~al.\/}, {\it Adv. Mater.\/} {\bf 25}, 1065 (2013).
	
	\bibitem{Kou2013}
	X.~Kou, {\it et~al.\/}, {\it ACS Nano\/} {\bf 7}, 9205 (2013).
	
	\bibitem{Po2018}
	H.~C. Po, L.~Zou, A.~Vishwanath, T.~Senthil, {\it Phys. Rev. X\/} {\bf 8},
	031089 (2018).
	
	\bibitem{Zou2018}
	L.~Zou, H.~C. Po, A.~Vishwanath, T.~Senthil, {\it Phys. Rev. B\/} {\bf 98},
	085435 (2018).
	
	\bibitem{Kang2018}
	J.~Kang, O.~Vafek, {\it Phys. Rev. X\/} {\bf 8}, 031088 (2018).
	
	\bibitem{Koshino2018}
	M.~Koshino, {\it et~al.\/}, {\it Phys. Rev. X\/} {\bf 8}, 031087 (2018).
	
	\bibitem{Amet2013}
	F.~Amet, J.~R. Williams, K.~Watanabe, T.~Taniguchi, D.~Goldhaber-Gordon, {\it
		Phys. Rev. Lett.\/} {\bf 110}, 216601 (2013).
	
	\bibitem{AmetThesis}
	F.~Amet, {\it \normalfont{thesis,}\/} {\bf \normalfont{Stanford University}}
	(2014).
	
	\bibitem{Moon2014}
	P.~Moon, M.~Koshino, {\it Phys. Rev. B\/} {\bf 90}, 155406 (2014).
	
	\bibitem{Mucha-Kruczynski2013}
	M.~Mucha-Kruczy{\'{n}}ski, J.~R. Wallbank, V.~I. Fal'Ko, {\it Phys. Rev. B\/}
	{\bf 88}, 205418 (2013).
	
	\bibitem{Jung2015}
	J.~Jung, A.~M. DaSilva, A.~H. MacDonald, S.~Adam, {\it Nat. Commun.\/} {\bf 6},
	6308 (2015).
	
	\bibitem{Zhang2018}
	Y.-H. {Zhang}, D.~{Mao}, Y.~{Cao}, P.~{Jarillo-Herrero}, T.~{Senthil}, {\it
		\normalfont{https://arxiv.org/abs/1805.08232}\/}  (2018).
	
	\bibitem{Esquinazi2003}
	P.~Esquinazi, {\it et~al.\/}, {\it Phys. Rev. Lett.\/} {\bf 91}, 227201 (2003).
	
	\bibitem{Mombru2005}
	A.~W. Mombr{\'{u}}, {\it et~al.\/}, {\it Phys. Rev. B\/} {\bf 71}, 100404
	(2005).
	
	\bibitem{Pardo2006}
	H.~Pardo, R.~Faccio, F.~M. Ara{\'{u}}jo-Moreira, O.~F. {De Lima}, A.~W.
	Mombr{\'{u}}, {\it Carbon\/} {\bf 44}, 565 (2006).
	
	\bibitem{Cervenka2009}
	J.~{\v{C}}ervenka, M.~I. Katsnelson, C.~F.~J. Flipse, {\it Nat. Phys.\/} {\bf
		5}, 840 (2009).
	
	\bibitem{Yeh1989}
	N.~C. Yeh, K.~Sugihara, M.~S. Dresselhaus, G.~Dresselhaus, {\it Phys. Rev. B\/}
	{\bf 40}, 622 (1989).
	
	\bibitem{Nicholls1990}
	J.~T. Nicholls, E.~J. McNiff, G.~Dresselhaus, {\it Phys. Rev. B\/} {\bf 42},
	5555 (1990).
	
	\bibitem{Gao2015}
	D.~Gao, {\it et~al.\/}, {\it J. Mater. Chem. C\/} {\bf 3}, 12230 (2015).
	
	\bibitem{Zhou2016}
	P.~Zhou, C.~Q. Sun, L.~Z. Sun, {\it Nano Lett.\/} {\bf 16}, 6325 (2016).
	
	\bibitem{Lian2018}
	B.~Lian, X.-Q. Sun, A.~Vaezi, X.-L. Qi, S.-C. Zhang, {\it Proc. Natl. Acad.
		Sci. U.S.A.\/} {\bf 115}, 10938 (2018).
	
	\bibitem{He2017}
	Q.~L. He, {\it et~al.\/}, {\it Science\/} {\bf 357}, 294 (2017).
	
	\bibitem{Mahoney2017}
	A.~C. Mahoney, {\it et~al.\/}, {\it Nat. Commun.\/} {\bf 8}, 1836 (2017).
	
	\bibitem{Bhandari2016}
	S.~Bhandari, {\it et~al.\/}, {\it Nano Letters\/} {\bf 16}, 1690 (2016).
	
	\bibitem{Wang2013b}
	L.~Wang, {\it et~al.\/}, {\it Science\/} {\bf 342}, 614 (2013).
	
	\bibitem{Zibrov2017}
	A.~A. Zibrov, {\it et~al.\/}, {\it Nature\/} {\bf 549}, 360 (2017).
	
	\bibitem{Kivelson1992}
	S.~Kivelson, D.-H. Lee, S.-C. Zhang, {\it Phys. Rev. B\/} {\bf 46}, 2223
	(1992).
	
	\bibitem{Hilke1998}
	{\it Nature\/} {\bf 395}, 675 (1998).
	
\end{thebibliography}
\end{document}